\documentclass[pra,groupedaddress,twocolumn,floatfig,showpacs]{revtex4-1}
\usepackage{graphicx,color}
\usepackage{amsmath,amssymb,bm}
\usepackage{braket}		% Dirac notation

\definecolor{darkred}{rgb}{0.90,0,0}
\definecolor{darkgreen}{rgb}{0,0.60,.2}
\definecolor{darkblue}{rgb}{0,0,1}
\definecolor{grey}{cmyk}{0,0,0,0.25}
\definecolor{orange}{cmyk}{0,0.6,0.8,0}

\begin{document}
\title{Relaxation and thermalization  in the  one-dimensional Bose-Hubbard model: A case study for the interaction quantum quench  from the atomic limit}

\author{S.\ Sorg}
\author{L.\ Vidmar}
\author{L.\ Pollet}
\author{F. Heidrich-Meisner}
\affiliation{Department of Physics and Arnold Sommerfeld Center for Theoretical Physics,
Ludwig-Maximilians-Universit\"at M\"unchen, D-80333 M\"unchen, Germany}

\begin{abstract}
Motivated by recent experiments, we study the relaxation dynamics and thermalization in the one-dimensional
Bose-Hubbard model induced by a global interaction quench. Specifically, we start from an initial state 
that has exactly one boson per site and is  the ground state of a system with infinitely strong repulsive interactions at unit filling.
Using exact diagonalization and the density matrix renormalization group method, we compute the time dependence
of such observables  as the multiple occupancy and the momentum distribution function. Typically,
the relaxation to stationary values occurs over just a few tunneling times. 
The stationary values are identical to the so-called diagonal ensemble on the system sizes accessible to our numerical methods 
and we further observe that the micro-canonical
ensemble describes the time averages of many observables reasonably well for small and intermediate interaction strength. 
The expectation values of observables in the 
canonical ensemble agree quantitatively  with  the time averages obtained from the quench at small interaction strengths,
and qualitatively provide a good description  
even in parameter regimes where the micro-canonical ensemble is not applicable due to finite-size effects.
We discuss our numerical results in the framework of the 
eigenstate thermalization hypothesis. Moreover, we also observe that the diagonal and the
canonical ensemble are practically identical for our initial conditions already on the level of 
their respective energy distributions for small interaction strengths. Finally, we discuss implications of our results for the interpretation of
a recent sudden expansion experiment [Phys. Rev. Lett. {\bf 110}, 205301 (2013)], in which the same interaction quench was  realized.
\end{abstract}

\maketitle

% 75.10.Pq  Magnetic ordering, general theory and models of spin chain models
% 71.27.+a  Strongly correlated electron systems
% 75.40.Mg  Computer modeling and simulation of magnetic critical points
% 05.60.Gg  Transport processes, quantum

\section{Introduction}
\label{sec:intro}

The non-equilibrium dynamics of closed many-body quantum systems is attracting considerable attention, 
fueled by both fundamental theoretical questions and a surge of experiments with ultra-cold quantum gases \cite{greiner02,kinoshita06,hofferberth07,trotzky12,gring12,cheneau12,schneider12,kuhnert13,langen13,meinert13,fukuhara13,ronzheimer13,braun14}
that give direct access to studying the real-time evolution of such systems.
As one of the primary goals for theory, one seeks to understand how steady states and thermal behavior emerge
in a closed quantum system \cite{polkovnikov11,rigol08,gemmerbook}. This case is typical for quantum gas experiments
which are well isolated from the environment.
Thermalization in such a situation can occur in the sense that subsystems equilibrate with each other, 
and can thus be probed by studying the relaxation dynamics and  thermal behavior of {\it local} observables.

A typical protocol to drive a many-body system out of equilibrium
is a quantum quench, in which one parameter of the Hamiltonian is changed instantaneously.
For such quantum quenches, in a large number of examples (see, e.g., \cite{kollath07,calabrese07,manmana07,cazalilla06,cramer08,barthel08,rossini09} and \cite{polkovnikov11,yukalov11} for a review), the emergence of thermal steady states has been
investigated and numerically demonstrated for {generic} models \cite{rigol08}, which do not possess any unusual conservation laws. 

In order to pinpoint  some of the questions addressed in this field, it is instructive to consider the expression for the
time evolution for a pure state, appropriate for a closed quantum system. Consider an observable $\hat A$ and its time evolution under the Hamiltonian $H$,
 which, for instance, could be induced by a quantum quench (we set $\hbar=1$):
\begin{equation}
\langle \hat A(t) \rangle = \langle \psi(t) |\hat A| \psi(t)\rangle 
\end{equation}
with 
\begin{equation}
|\psi(t) \rangle= e^{-iHt} |\psi(t=0)\rangle \,.
\end{equation}
In the eigenbasis $|\alpha\rangle$ of $H$ with $H|\alpha\rangle = E_\alpha | \alpha\rangle$, we can write:
\begin{equation}
 |\psi(t=0) \rangle =  \sum_\alpha c_\alpha |\alpha\rangle
\end{equation}
and therefore, neglecting degeneracies, 
\begin{equation}
\langle \hat A(t)\rangle = \sum_{\substack{\alpha\neq \alpha'}} A_{\alpha \alpha'} c_{\alpha}^* c_{\alpha'} e^{-i(E_{\alpha'}-E_{\alpha})t} + \sum_{\alpha} |c_\alpha|^2 A_{\alpha\alpha}\,,
\label{eq:a}
\end{equation}
with $A_{\alpha \alpha'} = \langle \alpha | \hat A|\alpha'\rangle$.
In a {generic} many-body system, one expects the first term in Eq.~\eqref{eq:a} to decay to zero due to dephasing \cite{rigol08,barthel08,khatami13,srednicki99} (see Ref.~\cite{banuls11} for a weaker definition of relaxation).
Following \cite{rigol08}, we introduce the term {\it diagonal ensemble} for the statistical mixture of states reproducing the infinite time average
\begin{equation}
\hat \rho_{\rm diag} = \sum_\alpha |c_\alpha|^2\,  \ket{\alpha}\bra{\alpha} \label{eq:diag}\,.
\end{equation}

One of the main questions is  whether
\begin{equation}
 \langle \hat A \rangle _{\rm diag}  = \sum_{\alpha} |c_\alpha|^2 A_{\alpha \alpha} = \langle \hat A \rangle _{\rm thermal} \; ,
\label{eq:gibbs}
\end{equation}
where $  \langle \hat A \rangle _{\rm thermal}$ is evaluated in a standard  thermal ensemble,
which implies independence of the initial state.

A second question pertains to the relaxation dynamics, namely how   $\langle \hat A(t)\rangle $
approaches its steady-state value. This could be  
an exponential, Gaussian, or a power-law approach \cite{barmettler09,cramer08,zangara13,torres-herrera14}, and related to that, the 
behavior and decay of temporal fluctuations  are of relevance to define an equilibrated state \cite{reimann08,linden09,khatami13}.
Equilibration may not necessarily occur in one step. For instance,
the possibility
of prethermalized metastable states has been demonstrated in theoretical studies \cite{berges04,moeckel08,eckstein08}.
Indications for prethermalization in experiments with one-dimensional (1D) Bose gases were reported on in Ref.~\cite{gring12}.

\begin{figure}[t]
        \centering
        \includegraphics[width=.9\columnwidth]{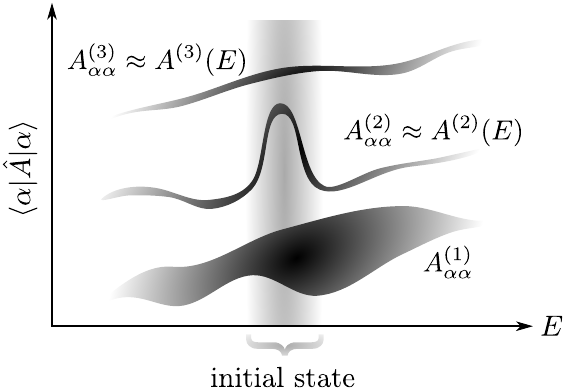}
        \caption{(Color online) Sketch of possible distributions of eigenstate expectation values of an observable $\hat A$. While $A^{(1)}_{\alpha\alpha}$ is broad with large fluctuations even between eigenstates close in energy, $A^{(2)}_{\alpha\alpha}$ and $A^{(3)}_{\alpha\alpha}$ can be considered sharp functions $A(E)$ of the energy, for which for a sufficiently large system and thus a narrow state, the ETH should hold.
On a finite system, where initial states can be so broad in energy (indicated by the shaded area) as to resolve
large variations of $A(E)$ with energy, $A^{(2)}_{\alpha\alpha}$ may lead to
non-thermal behavior. Since  $A^{(3)}_{\alpha\alpha}$ has a simple linear dependence on energy, the ETH is expected to apply in such
a case for sufficiently large systems.
        }\label{fig:sketch_ETH}
\end{figure}

Clearly, one desires criteria to be able to tell beforehand whether a given observable for a given model thermalizes or not.
One framework in which this  is addressed is the eigenstate thermalization hypothesis (ETH) \cite{srednicki94,deutsch91,rigol08}.
The ETH states that if the expectation values of an observable $\hat A$ in the eigenbasis of the post-quench Hamiltonian
are a smooth function of energy, then Eq.~\eqref{eq:gibbs} holds for the micro-canonical ensemble. 
In other words,  under the assumption
that the dependence of the $\braket{\alpha|\hat A|\alpha}$ on $E_\alpha$ can be approximated by a linear function over the range of the initial state,
the micro-canonical ensemble  describes steady-state values of observables in quantum quenches well. 
These concepts are meant to apply to large systems in the thermodynamic limit, while numerical
simulations and cold gas experiments work with finite  particle numbers.
Typical situations are illustrated in Fig.~\ref{fig:sketch_ETH}.

The validity of the ETH has been tested and established in numerous numerical examples (see, e.g., \cite{rigol08,rigol09,rigol09a,santos10,rigol10a,neuenhahn12,beugeling14,steinigeweg14, roux10}).
Finite-size effects are important to understand since many studies
are carried out numerically \cite{biroli10,rigol09,roux10}. 
A notable exception from the ETH picture are integrable 1D quantum systems where extra conservation laws typically lead
to non-thermal behavior \cite{rigol08,rigol07}. To account for that the generalized Gibbs ensemble with infinitely many Lagrange parameters
was introduced in Ref.~\cite{rigol07} and studied for many integrable systems 
(see, e.g., \cite{rigol06,rigol07,barthel08,kollar08,essler12,konik12,iyer12,karrasch12a,fagotti13}).

 Many studies have emphasized the presence of  chaotic properties in the energy spectrum of many-body systems (see, e.g., \cite{flambaum97,santos10,rigol10a,santos12,santos12a}) for the emergenve of thermalization.
Such considerations give a handle on understanding the width and typical form of the energy distribution defined by an initial state,
which is essentially the product of the density of states with the weights given in Eq.~\eqref{eq:diag}.
We will apply these concepts to understand why the canonical ensemble works well in describing time averages of observables for our
quench.
For further discussions of thermalization alternative to the ETH and the connection between ETH and chaotic systems,  see, e.g., Refs.~\cite{santos10,biroli10,rigol12,mierzejewski13,sirker14}.

One of the many-body lattice systems that is  well established in quantum gas experiments is the Bose-Hubbard model (BHM) \cite{jaksch98,greiner02a},
for which a  number of non-equilibrium experiments have been performed \cite{greiner02,chen11,cheneau12,trotzky12,meinert13,ronzheimer13}.
This is also a model, for which a direct comparison between experimental results and theory is feasible \cite{trotzky12,ronzheimer13}.

In this work we will be interested in its one-dimensional version, for which the Hamiltonian is given as:
\begin{align}
H = -J \sum_{i=1}^{L} (a_i^\dag a_{i+1} + {\rm H.c.}) + \frac{U}{2} \sum_{i=1}^L {\hat n}_i ({\hat n}_i - 1)\,.
\end{align}
The operator $a^{\dag}_i$ creates a boson at site $i$, $\hat n_i = a^{\dag}_i a_i$ is the particle number operator and $L$ is the number of sites. We assume periodic boundary conditions (PBC), unless stated otherwise, and a lattice spacing of $a=1$. This system has a Mott-insulating (MI) phase at large $U/J$ at any integer filling 
and a superfluid (SF) phase in all other parameter regimes. At unit filling, the transition occurs at $(U/J)_{\rm crit }\approx 3.3$ \cite{Prokofev_etal_98,kuhner00}.

Various non-equilibrium topics have been investigated theoretically for the one-dimensional Bose-Hubbard model, including interaction quenches 
\cite{schuetzhold06,kollath07,fischer08,roux09,roux10a,rigol10,biroli10,roux10,cramer08,lux13,barmettler13,queisser14},
the relaxation dynamics starting from an ideal charge density wave state \cite{cramer08a,flesch08},  the propagation of correlations \cite{laeuchli08,barmettler12,natu13,carleo14}, the growth of entanglement in quantum quenches \cite{laeuchli08}, slow quantum quenches \cite{bernier11,bernier12},
 wave-packet propagation \cite{kollath05,kleine08},  sudden expansion dynamics \cite{rigol04,rigol05,collura12,ronzheimer13,rodriguez06,vidmar13},  quenches in two-component Bose gases at finite temperatures \cite{zhang12} and dynamics in trapped gases \cite{natu11,bernier11,bernier12}.
We mention that the question of thermalization has also been studied for the Bose-Hubbard model with dissipation, spontaneous emission,  or other couplings to the environment (see e.g., 
\cite{poletti12,schachenmayer14}).

In recent experiments, the non-equilibrium dynamics starting
from {\it product} states in real space has been studied, including  an ideal density-wave state \cite{trotzky12}
or the ground state of the BHM at $U/J=\infty$ and $n = N/L = 1$ (where $N$ represents the number of particles) \cite{cheneau12,ronzheimer13}
\begin{equation}
|\psi(t=0)\rangle = \prod_{i=1}^{L} a_i^{\dagger} | 0\rangle\,.
\label{eq:init0}
\end{equation}
The ensuing time evolution is then monitored as a function of the post-quench interaction strength.

Our work is motivated, in particular, by the experimental and numerical results from Ref.~\cite{ronzheimer13} for the one-dimensional case. In that study, the dynamics were in fact induced by
both quenching $J$ from $0$ to $J>0$, 
and a simultaneous removal of the trapping potential.
The quench of $J$ is, for the initial state given in Eq.~\eqref{eq:init0}, equivalent to
an interaction quench from $U/J=\infty$ to finite values.
A main result  of Ref.~\cite{ronzheimer13} is the observation of ballistic dynamics of strongly interacting
bosons in one dimension due to the integrability of hard-core bosons and their exact mapping to non-interacting fermions, in sharp contrast to the behavior
in two dimensions.  The experimental and numerical data also clearly unveiled a
 fast local relaxation that is relevant to understand the dynamics in particular at intermediate $U/J$.
This local relaxation is due to the interaction quench performed in the experiment.
As a notable consequence of the interaction quench, interacting bosons with, e.g.,  $U/J\sim 4> (U/J)_{\rm crit}$ expand ballistically if they start from the ground state in the
trap but for the expansion from Eq.~\eqref{eq:init0}, the interaction quench causes much slower, presumably diffusion-like dynamics \cite{ronzheimer13,vidmar13}.

The goal of our study is to understand relaxation and steady states for the specific initial state Eq.~(\ref{eq:init0}) of Refs.~\citep{ronzheimer13, cheneau12}. In particular, we intend to clarify whether experimentally measurable observables can, in the regime of intermediate $U/J\sim 4$,
 be described by standard thermal ensembles. The establishment of at least local equilibrium in these experiments can be seen as 
a prerequisite for diffusive transport.

We will use exact diagonalization (ED) and the time-dependent density matrix renormalization group (tDMRG) method \cite{daley04,white04,vidal04} to compute the time evolution of these
observables, extending first results from Ref.~\cite{ronzheimer13} (see also closely related recent work \cite{queisser14,barmettler13,lux13}).  We observe that steady-state values are approached very fast 
for all $U/J$. We then use exact diagonalization and quantum Monte Carlo simulations to compare the time averages of observables to various thermal
ensembles.

As a main result, we observe that the micro-canonical ensemble describes well many observables in the regime $U/J < 5$, while we 
  also identify examples where it fails on small systems in this parameter regime.
Moreover, we will discuss the results in the framework of the ETH and demonstrate that failures of the micro-canonical ensemble
in describing time averages can be traced back to simple finite-size effects, namely either 
the initial state being  too broad in energy or the total energy set by the initial condition sitting in the first gap in the large $U/J$ regime.
We estimate that even with typical particle numbers used in quantum gas experiments that are larger than what we can access numerically, thermal behavior in the sense of the ETH and of ensemble equivalence \cite{roux10a,ll} cannot be seen
for large $U/J\gg 4$ for the initial conditions considered here.

Surprisingly, the canonical ensemble describes the time averages of observables very well for small $U/J\lesssim 5$
and yields a qualitative good approximation  
for most observables studied here for all $U/J$, even where the micro-canonical ensemble is ill-defined due to finite-size effects.
In the small $U/J$ regime, this can be traced back to a remarkable agreement of ensemble energy distributions $\rho_\text{ens}(E) = p_\text{ens}(E) g(E)$, where $p_\text{ens}(E)$ is the occupation of eigenstates at energy $E$ and where $g(E)$ denotes the density of states. For the initial condition given in Eq.~(\ref{eq:init0}), the diagonal and the canonical energy distributions are practically identical at small $U/J$, an agreement, which gradually deteriorates as $U/J$ increases.
A similar observation for initial product states has been
reported  in Ref.~\cite{rigol11}.
While the focus of Ref.~\cite{rigol11}  was specifically on understanding thermalization in 
integrable systems, we emphasize that initial real-space product states can lead to time averages 
similar to canonical expectation values in the non-integrable case as well.

Our results for the density of states and the energy distribution are corroborated by studying the time evolution of the fidelity $F=\left|\langle\psi(t=0)| \psi(t) \rangle \right |^2$, i.e., the probability to recover the initial state at a given time $t$. The fidelity  is  related
 to the energy
distribution in Eq.~\eqref{eq:diag_en_distr} by a Fourier transformation. 
We observe a Gaussian decay of the fidelity for $0<U/J\lesssim 4$, with the time scale being directly related to the
width of the energy distribution $\rho_{\rm diag}(E)$. Qualitatively similar observations on the short-time dynamics of the fidelity
have been recently reported on in Refs.~\cite{torres-herrera14,torres-herrera14a}. 

Finally, we discuss implications of our results for the interpretation of the sudden expansion  dynamics in the regime of large $U/J \gg 4$, namely
we derive an analytical prediction for the expansion velocity (see \cite{ronzheimer13}) based on a two-component picture of single bosons and 
inert doublons.

The plan of this exposition is the following. 
Some technical aspects and definitions as well as the  numerical methods used in this work are briefly introduced in Sec.~\ref{sec:setup}.
Our main results for relaxation and thermalization in the  interaction quench are contained in Sec.~\ref{sec:quench}.
We first discuss the relaxation dynamics in the  
interaction quench in Sec.~\ref{sec:relax}
and then devote Sec.~\ref{sec:thermo} to the  comparison of  the time averages to the expectations from the 
micro- and the canonical ensemble. Section~\ref{sec:eth} contains a discussion of the ETH for our example, while we provide a detailed discussion on the validity of the 
micro- and the canonical ensemble in Secs.~\ref{sec:mc} and \ref{sec:can}, respectively.
The implications of our
results for the sudden expansion studied in Ref.~\onlinecite{ronzheimer13} are  presented in Sec.~\ref{sec:exp}.
We conclude with a summary of our results, Sec.~\ref{sec:sum}.
In  Appendix~\ref{sec:fidel}, we provide  a discussion of the short-time dynamics of the fidelity and its connection
to properties of the energy distribution.
Appendix \ref{sec:error} contains examples for the error analysis of time-dependent DMRG data.

%%%%%%%%%%%%%%%%%%%%%%%%%%%%%%%%%%%%%%%%%%%%%%%%%%%%%%%%%%%%%%%%%%%%%%%%%%%%%%%
\section{Set-up and definitions}
\label{sec:setup}

\subsection{Initial state}
\label{sec:state}
For the initial state given in Eq.~\eqref{eq:init0}, the total energy $E=0$ is independent of $U$, while the quench energy $\delta E$, i.e., 
the energy of the system with respect to the post-quench ground-state energy $E_0$, is 
\begin{equation}
\delta E = E- E_0 = |E_0|\,.
\end{equation}
The width of the initial state is 
\begin{equation}
\sigma_{\rm diag} = \sqrt{\langle H^2\rangle - \langle H \rangle^2}= {2} J\sqrt{L}\,\label{eq:sigma_width} 
\end{equation}
by a straightforward calculation (see Appendix \ref{sec:fidel}).
The energy distribution associated with the diagonal ensemble, Eq.~\eqref{eq:diag},
is defined as
\begin{equation}
\rho_{\rm diag}(E) =  \sum_{\alpha} |c_{\alpha}|^2 \delta(E-E_{\alpha})\,.
\label{eq:diag_en_distr}
\end{equation}

\subsection{Observables}
\label{sec:observables}
We will concentrate on  the fraction of atoms on multiply occupied sites and  the quasi-momentum distribution function (MDF).
First, the fraction $\nu_h$ of atoms on multiply occupied sites is given by
\begin{align}
	\nu_h := \braket{\hat \nu_h} := \frac{1}{L} \sum_{i=1}^L \sum_{m=2}^{N} m \braket{\hat{n}^{(m)}_i}
\end{align}
where $\hat{n}^{(m)}_i$ is defined through its action on the position basis Fock states, $\hat{n}^{(m)}_i \ket{..., n_i, ...} = \delta_{m, n_i} \ket{..., n_i, ...}$, yielding one if the occupancy on site $i$ is $m$ and zero otherwise.

Second, we compute the quasi-momentum distribution function $n_k$, given by
\begin{align}
		 n_k = \langle \hat{n}_k \rangle= \frac{1}{L}\sum_{l,m} e^{-ik(l-m)a} \braket{a^{\dag}_l a_m} \ ,
\end{align}
which we use for PBC.
The MDF is not a local observable but is one of the most frequently measured ones in experiments.
%For large quasi-momenta $k$, one can argue that the MDF probes short distance correlations.

\subsection{Statistical ensembles}
 
{\it Micro-canonical ensemble.}
The micro-canonical ensemble (denoted with a subscript mc throughout the paper)   consists of the  energy eigenstates at the system's  energy $E$, 
each occurring with the same probability.
Since we are dealing with a finite system and thus a discrete energy spectrum, we use the statistical mixture assigning
the same weight to all energy eigenstates with energies in a window $[E-\Delta E, E+\Delta E]$ around the system's energy $E$,
\begin{align}
	\hat \rho_{\text{mc}} &= \frac{1}{\mathcal{N}_{E, \Delta E}} \sum_{|E-E_\alpha| < \Delta E} \ket{\alpha} \bra{\alpha} \ ,
\end{align}
where $\mathcal{N}_{E, \Delta E}$ is the number of energy eigenstates in $[E-\Delta E, E+\Delta E]$.
In practice, we determine $\Delta E$ by plotting $\langle \hat A \rangle_{\rm mc}$ versus $\Delta E$
and choosing
$\Delta E$ such that  $\langle \hat A \rangle_{\rm mc}$ is independent of the value of $\Delta E$.
For consistency, we require that 
\begin{equation}
E = \langle H \rangle_{\rm mc}\,.
\end{equation}

{\it Canonical ensemble.}
The canonical ensemble (subscript can) is usually used to describe a system with  fixed particle number in thermal equilibrium with a heat bath at some temperature $T$.
It is the statistical mixture of states given by the density matrix
\begin{align}
	\hat \rho_\text{can} = \frac{1}{Z} e^{-\beta H} = \frac{1}{Z} \sum_{\alpha} e^{-\beta E_\alpha} \ket{\alpha}\bra{\alpha}
\end{align}
where $Z$ is the partition function $Z = \text{tr} (e^{-\beta H}) = \sum_{\alpha} e^{-\beta E_\alpha}$ and $\beta = 1/T_{\rm can}$ is the inverse  temperature (in units of 1/energy). In order to consistently describe the steady state of our pure system with a canonical ensemble, $\beta$ will be chosen such that the ensemble energy expectation value reproduces the system's energy $E$,
\begin{align}
	\braket{H}_{\rm can} = \bra{\psi(t)} H \ket{\psi(t)} = E =0\ .
\end{align}
The energy distribution associated with the canonical ensemble 
is defined as
\begin{equation}
\rho_{\rm can}(E) = \frac{1}{Z}   \sum_{\alpha}    e^{-\beta E_{\alpha}} \delta(E-E_{\alpha})\,.
\label{eq:can_en_distr}
\end{equation}

{\it Grand-canonical ensemble.}
We will also compute the temperatures obtained in the grand-canonical ensemble (subscript gc)
\begin{equation}
\hat \rho_{\rm gc} = \frac{1}{Z} e^{-\beta (H-\mu \hat N)},
\end{equation}
where we determine $\beta=1/T_{\rm gc}$ and the chemical potential $\mu$ such that they fulfill the equations
\begin{equation}
E= \mbox{tr}\lbrack H \hat \rho_{\rm gc} \rbrack; \quad N=  \mbox{tr}\lbrack \hat N \hat \rho_{\rm gc} \rbrack\,.
\end{equation}

\subsection{Numerical methods}
\label{sec:num}

For ED simulations, we use a Krylov-space based time-evolution scheme \cite{manmana05} to propagate the wave function, which, using ED methods,
is possible for $L\sim 14$. We do not impose  a cut-off $N_{\rm cut}$ in the local number of bosons.
We obtain the time evolution of several observables using this method, while results
for larger systems with $L\sim 20$ are obtained from tDMRG simulations. We use a Trotter-Suzuki method
to propagate the wave-function using tDMRG; the time step is typically chosen to be $\delta t\sim 0.02/J$. The discarded
weight which controls the accuracy of the truncation involved in tDMRG \cite{schollwoeck11,schollwoeck05} is varied between
$\delta \rho \sim 10^{-4}$ and $\delta \rho \sim 10^{-7}$. 
Examples demonstrating the numerical quality of the  time-dependent DMRG data are shown in Appendix~\ref{sec:error}.
The cut-off $N_{\rm cut}$ was varied between $ 4<N_{\rm cut} <9 $ 
to ensure convergence.
We perform tDMRG runs  for open boundary conditions (OBC), while ED results are for PBC, unless stated otherwise. 

Exact diagonalization allows us to obtain the complete spectrum for systems with $L\sim 10$ without any truncation in the 
local number of bosons (i.e., $N_{\rm cut}=N$) by exploiting translational invariance and conservation of total particle number.
This information is used to construct thermal ensembles for comparison with the time averages of observables. For large systems, 
we use quantum Monte Carlo simulations to compute thermal expectation values.
Explicitly, we use path integral Monte Carlo with worm updates \cite{Prokofev_etal_98}, in the implementation of~\cite{Pollet07}. For a recent review, see~\cite{Pollet12}.
If not stated otherwise, then PBC are imposed in the QMC simulations.

\begin{figure}[!t]
\includegraphics[width=\columnwidth]{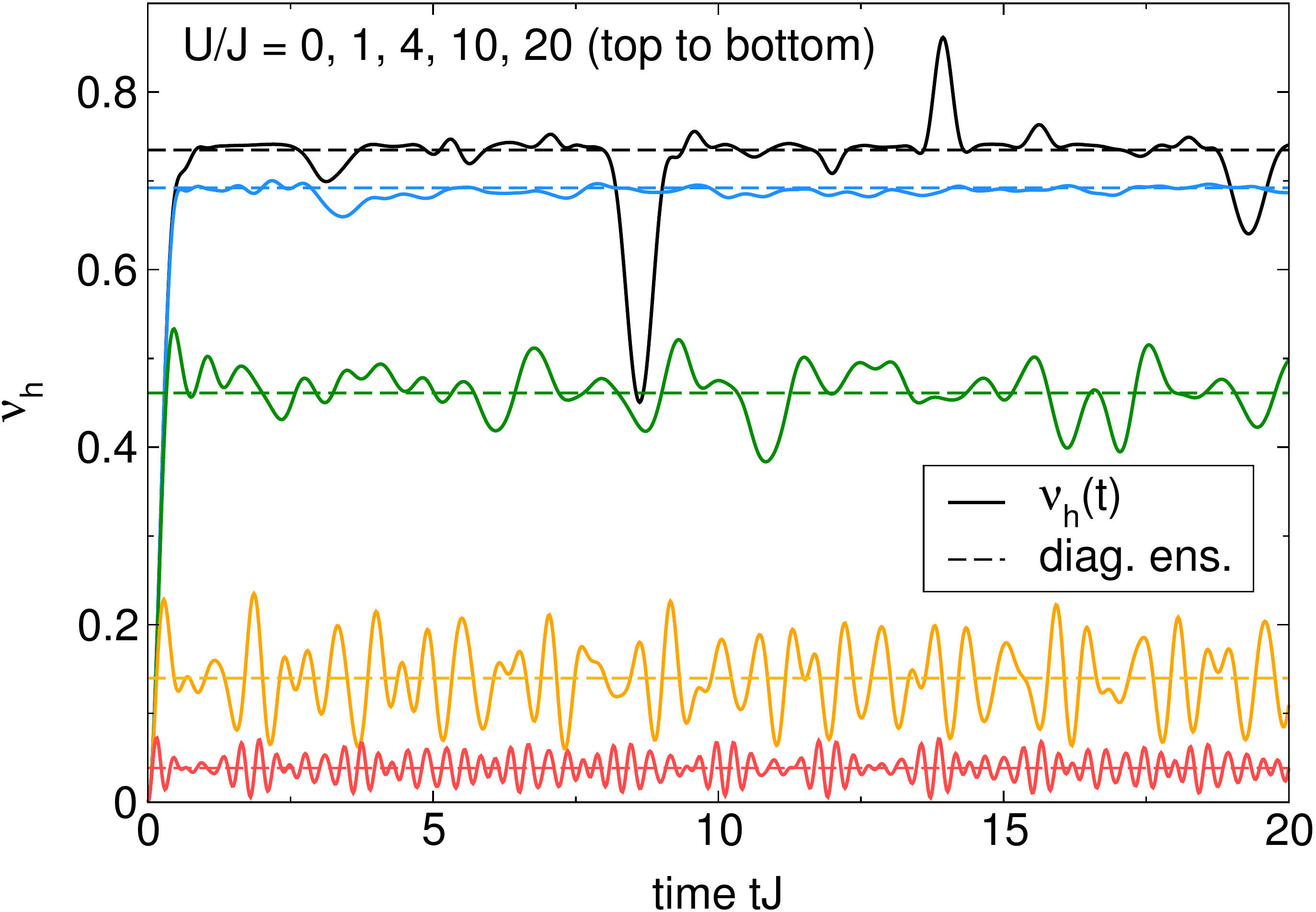}
\caption{(Color online)
Time evolution (solid lines) of the fraction of atoms on multiply occupied sites $\nu_h$ for the  quench to  $U/J=0,1,4,10,20$ (top to bottom),  $L=10$, PBC.
Dashed lines: Diagonal ensemble.
Data for $tJ<2$ was shown in Ref.~\onlinecite{ronzheimer13}.
}\label{fig:nh}
\end{figure}

%%%%%%%%%%%%%%%%%%%%%%%%%%%%%%%%%%%%%%%%%%%%%%%%%%%%%%%%%%%%%%%%%%%%%%%%%%%%%%%
\section{Interaction quench}
\label{sec:quench}
\subsection{Relaxation dynamics}
\label{sec:relax}

In this section, we present our results for the time-evolution of various observables, starting from the initial conditions
given by Eq.~\eqref{eq:init0}. 
The filling is consequently always $n=N/L=1$.

\begin{figure}[!t]
\includegraphics[width=.96\columnwidth]{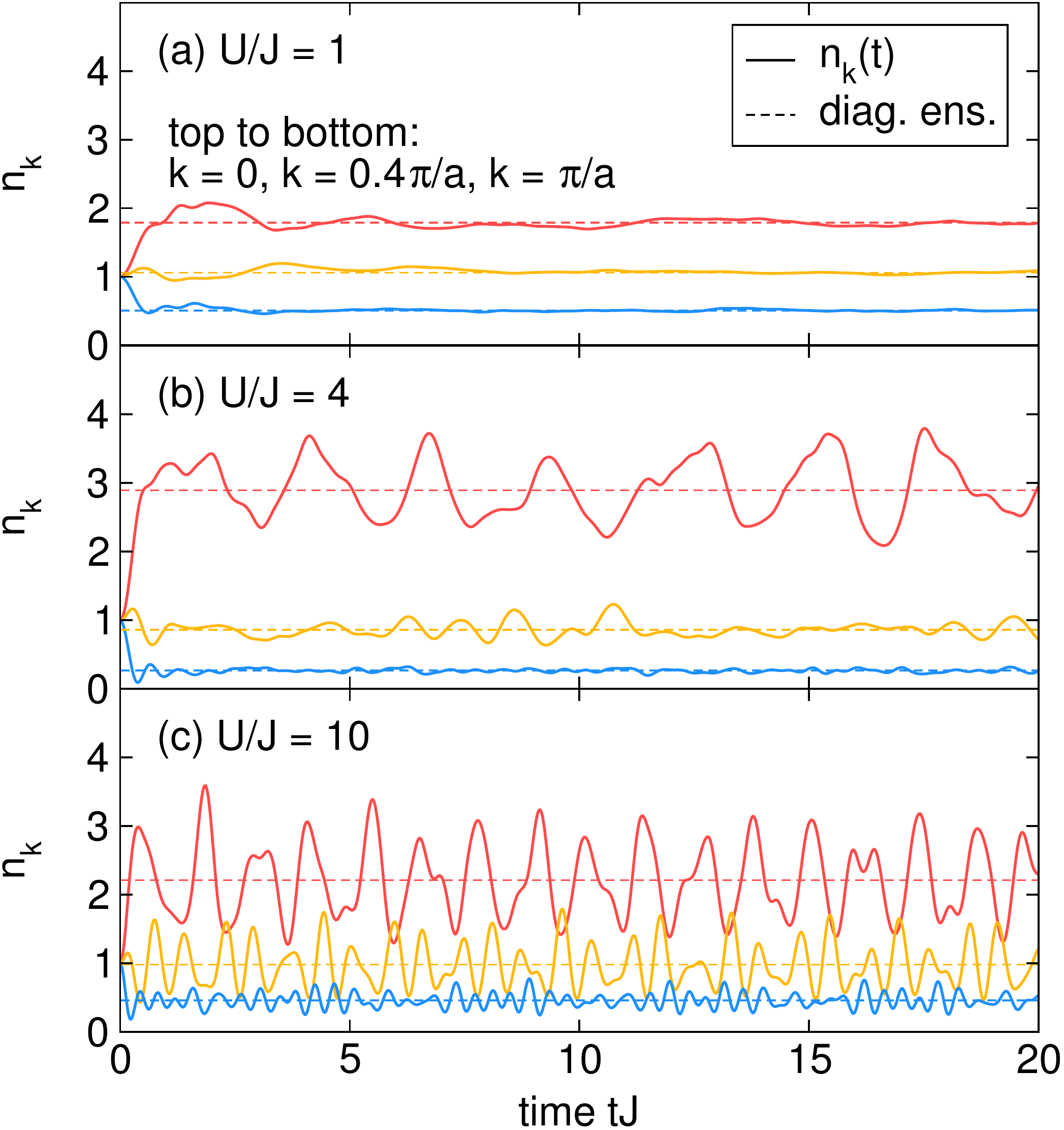}
\caption{(Color online)
Time evolution (solid lines) and diagonal ensemble (dashed lines) of $n_k$, $ka=0, 0.4\pi, \pi$, after the  quench to finite $U/J<\infty$ 
for  $L=10$ and PBC:
 (a) $U=J$; (b) $U=4J$; (c) $U=10J$.
}\label{fig:nk1}
\end{figure}

Typical results for the time evolution of $\nu_h$ and the MDF $n_k$ are shown in Figs.~\ref{fig:nh} and \ref{fig:nk1}.
Evidently, $\nu_h$ quickly increases from zero for all values of $U/J$ 
and approaches its stationary value after $tJ \sim 2$, as was already discussed in Ref.~\onlinecite{ronzheimer13}.
At $U=0$, there are  large deviations from the time average at times $tJ=8,14,19$, which are finite-size effects as their
amplitude decreases and their position moves to larger times as $L$ increases.

The behavior for $n_k$ is quite similar. We display results for three different quasi-momenta in Fig.~\ref{fig:nk1}.
In all cases, also including density-density correlations (not shown here), the time average is identical to the expectation value in the diagonal ensemble defined in Eq.~\eqref{eq:diag}. 
Note that at $U=0$, analytical results for simple observables can be obtained, see, e.g., Ref.~\cite{queisser14}.

\begin{figure}[!t]
\includegraphics[width=0.96\columnwidth]{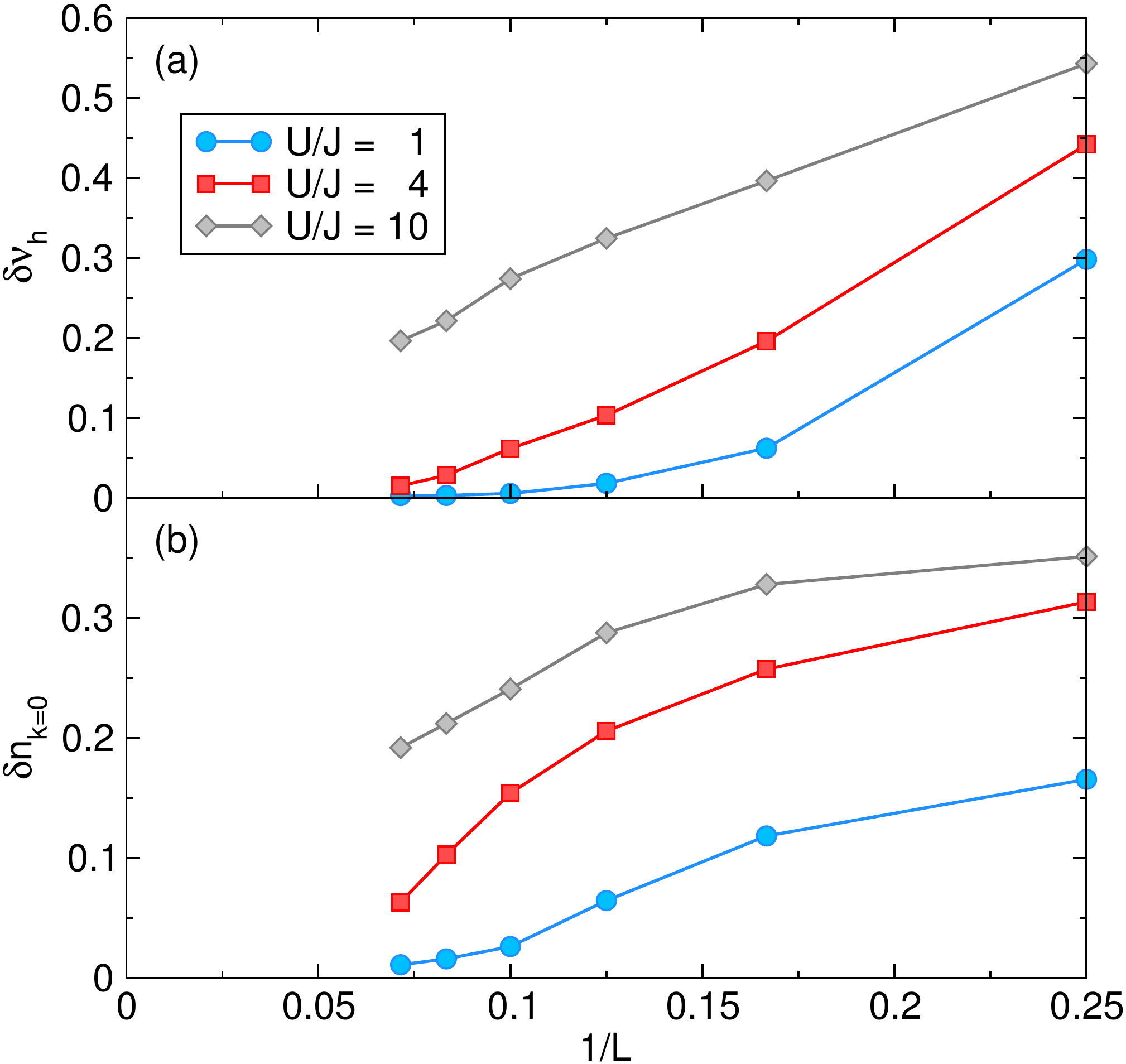}
\caption{(Color online)
	Relative temporal standard deviation, Eq.~(\ref{deltaA}),
	of (a) $\nu_h$ and (b) $n_{k=0}$ as a function of system size $L\leq 14$ after the quench from $U/J=\infty$ to $U/J=1,4,10$ for the time interval $tJ \in [5, 20]$ (unit filling $N/L=1$, PBC). 
	In all cases, the fluctuations quickly decrease with system size. 
}\label{fig:nk3}
\end{figure}

The oscillations around the time average are typically small
for  $U/J\sim 1$
but larger for large $U/J$ 
(compare Figs.~\ref{fig:nh} and \ref{fig:nk1}).
This is simply related to the number of accessible eigenstates in the vicinity of the
 total energy $E=0$.
As expected \cite{cramer08,khatami13,srednicki99,rigol08,zangara13}, these temporal fluctuations around the mean value quickly decrease with system size, as is evident from our data
shown in Fig.~\ref{fig:nk3}, where we show the relative standard deviation for $\nu_h$ and $n_{k=0}$,
defined by
\begin{equation} \label{deltaA}
\delta A = \frac{\sigma_{A,t}}{\overline{A}_t} \; ,
\end{equation}
where 
\begin{equation}
\sigma_{A,t}^2 = \frac{1}{t_2-t_1} \int_{t_1}^{t_2} \langle \hat A(t)  \rangle ^2 dt - \, \overline{A}_t^2
\end{equation}
and 
\begin{equation}
\overline{A}_t =  \frac{1}{t_2-t_1} \int_{t_1}^{t_2} \langle \hat A(t)  \rangle dt  \,.
\end{equation}
Evidently, in all cases, the fluctuations decay to zero with system size, in some cases consistent with an exponential dependence on $L$.

%%%%%%%%%%%%%%%%%%%%%%%%%%%%%%%%%%%%%%%%%%%%%%%%%%%%%%%%%%%%%%%%%%%%%%%%%%%%%%%
\subsection{Comparison of time averages with statistical ensembles}
\label{sec:thermo}

\begin{figure}[!t]
\includegraphics[width= 0.96\columnwidth]{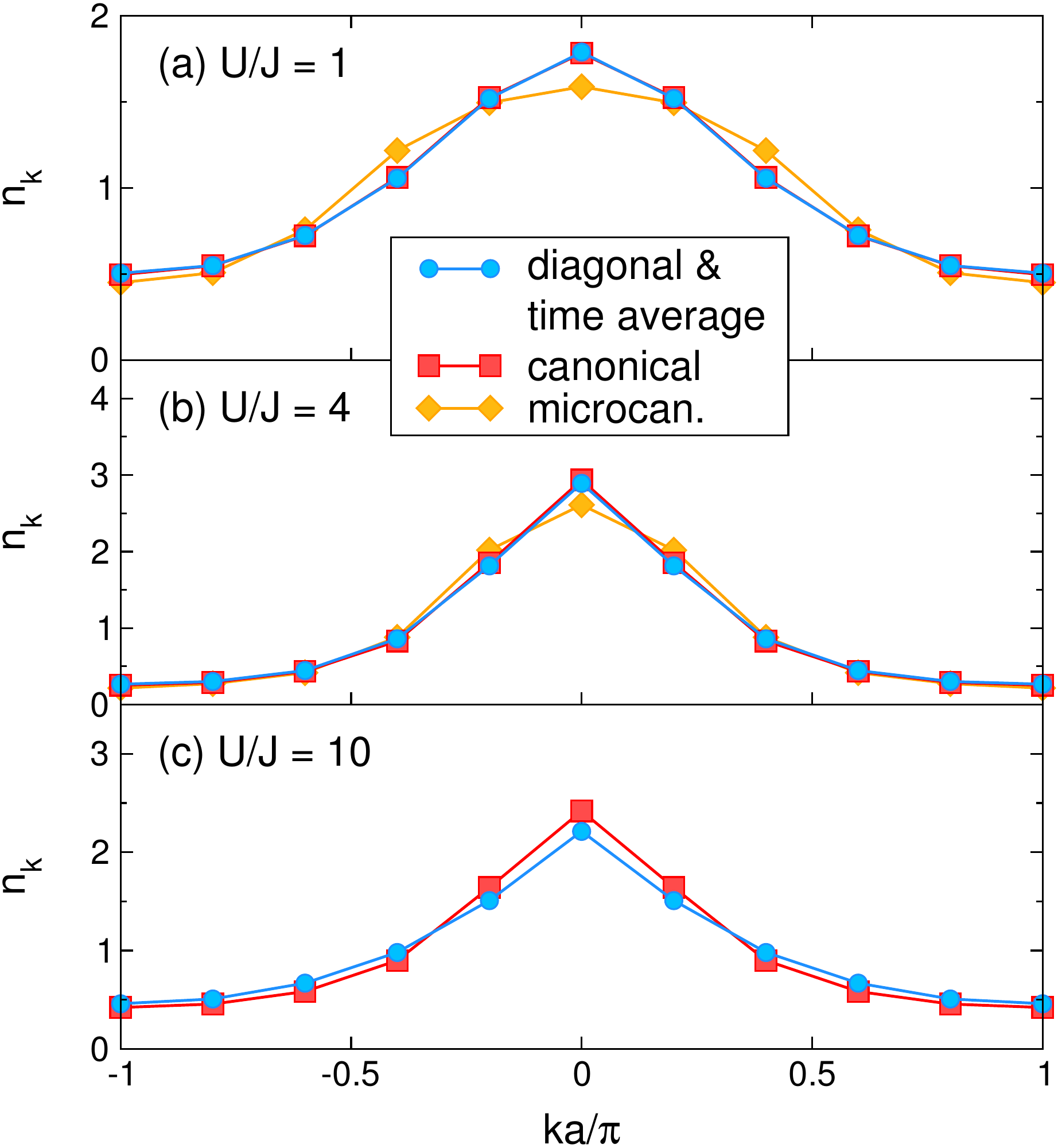}
\caption{(Color online) MDF
$n_k$ versus $k$: diagonal (circles), canonical (squares) and micro-canonical (diamonds) ensembles for a system with $L=10$ and PBC. (a) $U=J$; (b) $U=4J$ (c) $U=10J$. For this system size, the micro-canonical ensemble cannot properly be defined for $U/J=10$. 
}\label{fig:nk2}
\end{figure}

\begin{figure}[!t]
\includegraphics[width=0.96\columnwidth]{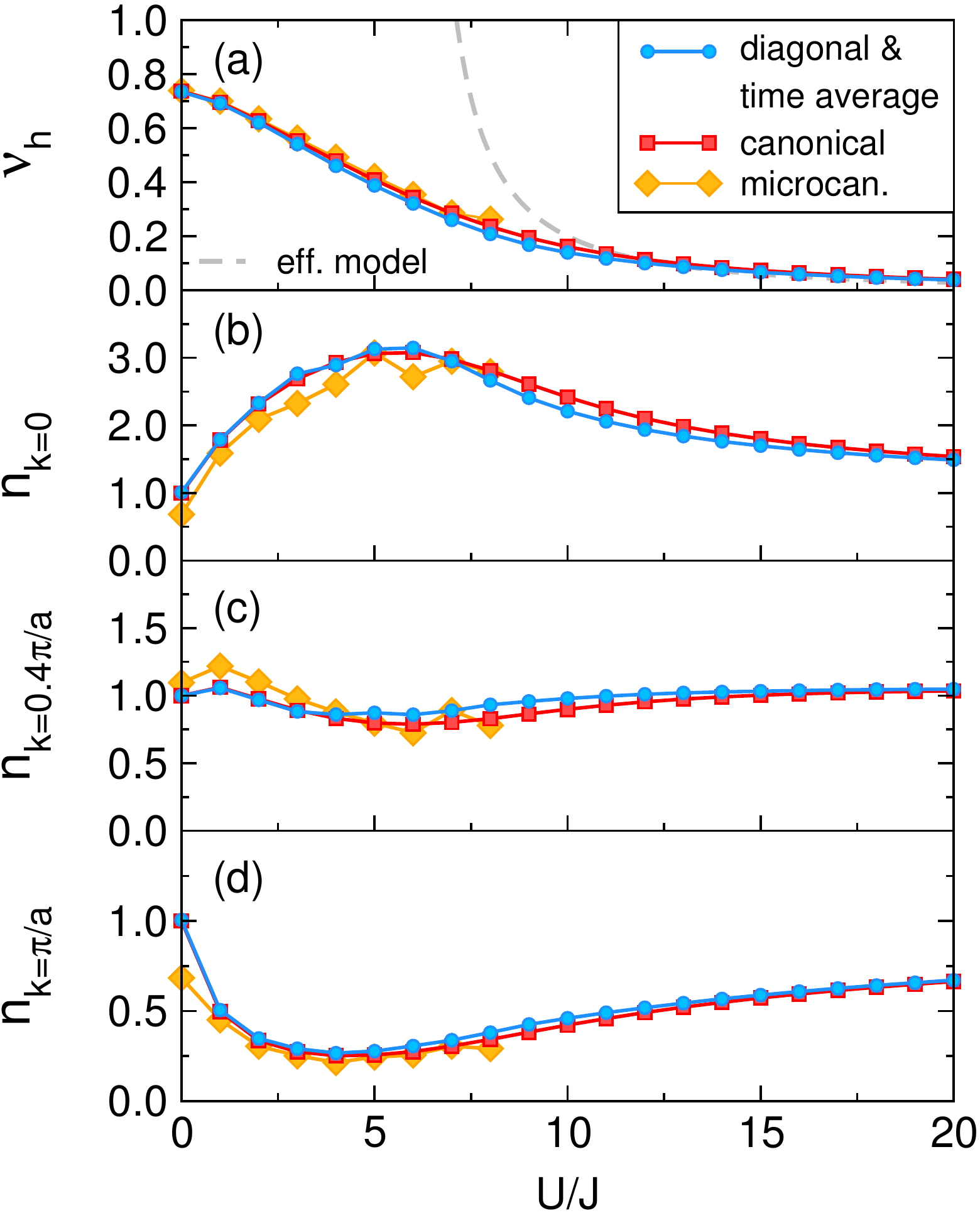}
\caption{(Color online)
Comparison of the diagonal with the canonical and micro-canonical ensembles as a function of $U/J$ for a system with $L=10$ and PBC after the global interaction quench from infinite to finite $U/J$.
{(a)} $\nu_h$, and $n_k$ for {(b)} $k=0$, {(c)} $k=0.4\pi/a$, {(d)} $k=\pi/a$.
The dashed line in (a) represents the fraction of doublons, Eq.~(\ref{doublons_diag}), in the diagonal ensemble of an effective two-level system discussed in Sec.~\ref{subsec:fraction}.
}
\label{fig:tav}
\end{figure}

We now turn our attention to the comparison of the time averages of observables and the thermal expectation values calculated in different ensembles. 
In Figs.~\ref{fig:nk2} and \ref{fig:tav}, we show results for $n_k$ and $\nu_h$,  plotting
the diagonal ensemble (identical to the time average), the micro-canonical ensemble, and the canonical ensemble.

First of all, let us stress that we show results for the micro-canonical ensemble only if (i) it correctly produces the
right average energy equal to the total energy defined by the quench and (ii) if a range of $\Delta E$ can be found over which the micro-canonical expectation values
are independent of $\Delta E$.
These criteria can be met  for $0<U/J \lesssim 8$. We find that the micro-canonical ensemble systematically describes
well $\nu_h$ (see Fig.~\ref{fig:tav}(a)) and density correlations in that regime (results not shown here).

In the case of the MDF, $n_{k=0}$ is typically not well described by the micro-canonical ensemble (see Fig.~\ref{fig:tav}(b)). 
Since $n_{k=0}$ is typically underestimated in the micro-canonical ensemble compared to the diagonal ensemble,
for other values of $k\neq 0$, $n_k$ is too large in the micro-canonical ensemble because of the normalization of the MDF.
Indeed, for some finite quasi-momenta $k>0$, there are large deviations as exemplified in the case $U/J=1$ and $k=0.4\pi/a$ shown in Fig.~\ref{fig:nk2}(a). 

The canonical ensemble describes the diagonal ensemble very well for $U/J\lesssim 5$ (see Figs.~\ref{fig:nk2}(a),~\ref{fig:nk2}(b), and ~\ref{fig:tav}), even at $U=0$. Surprisingly, even where the micro-canonical ensemble fails on small systems, i.e., for $U/J\gtrsim 8$,
the canonical ensemble still qualitatively reproduces the $U/J$ dependence of all quantities and the quasi-momentum dependence
of the MDF
quite well, while quantitative differences between the canonical and diagonal ensemble are typically the largest  for $U/J \approx 10$.

\subsection{Discussion of the ETH}
\label{sec:eth}
In order to understand why the thermal ensembles work in some cases and fail in other examples, it is instructive to revisit the question of the validity
of the ETH for this model (see Refs.~\cite{roux10,biroli10,beugeling14}), focusing on the regime in energy defined by our initial condition,
i.e., $ -\sigma_{\rm diag}< E < \sigma_{\rm diag}$, recalling that the total energy is $E=0$, independently of $U/J$. 

\begin{figure}[!t]
        \includegraphics[width=0.99\columnwidth]{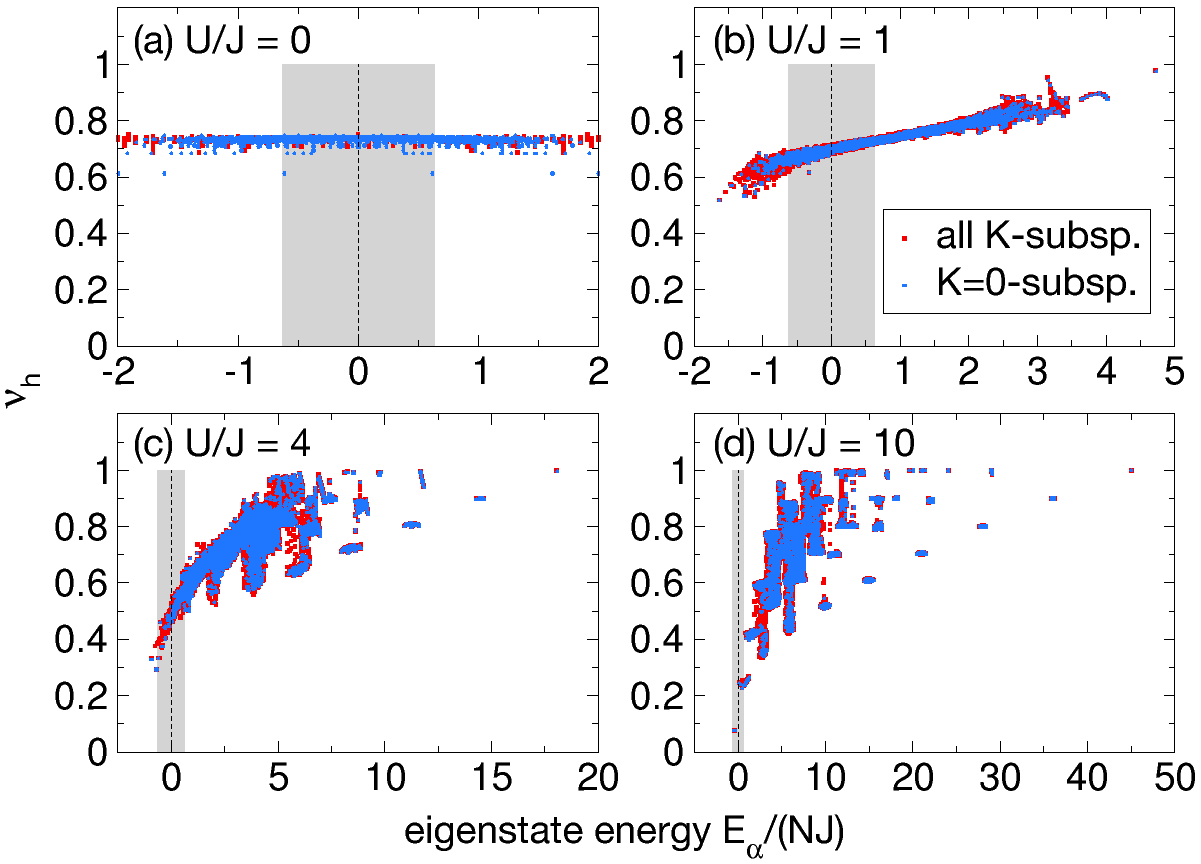}
        \caption{(Color online) Eigenstate expectation values $\bra{\alpha}\hat{\nu}_h\ket{\alpha}$ of $\nu_h$ for (a) $U=0$, (b) $U=J$, (c) $U=4J$, (d) $U=10J$, and $N=L=10$.
Red (dark): All energy eigenstate expectation values. Blue (light): Expectation values in energy eigenstates in the $K$=0 subspace, which are  the only ones relevant for the diagonal ensemble for our initial condition. Vertical lines mark the total energy in our quench, $E=0$. 
For the integrable case $U/J=0$, degenerate energy eigenvalues exist in each $K$-subspace, such
 that the choice of the basis is not unique. Here and in Figs.~\ref{fig:mel2}-\ref{fig:mel4} we represent the diagonal elements of observables in the simultaneous
eigenbasis of $n_k$.  Shaded area: Width of the initial state, $\sigma_{\rm diag}/(NJ) = 2/\sqrt{N}\approx 0.63 $. 
       } \label{fig:mel1}
\end{figure}

\begin{figure}[!t]
        \includegraphics[width=0.49\textwidth]{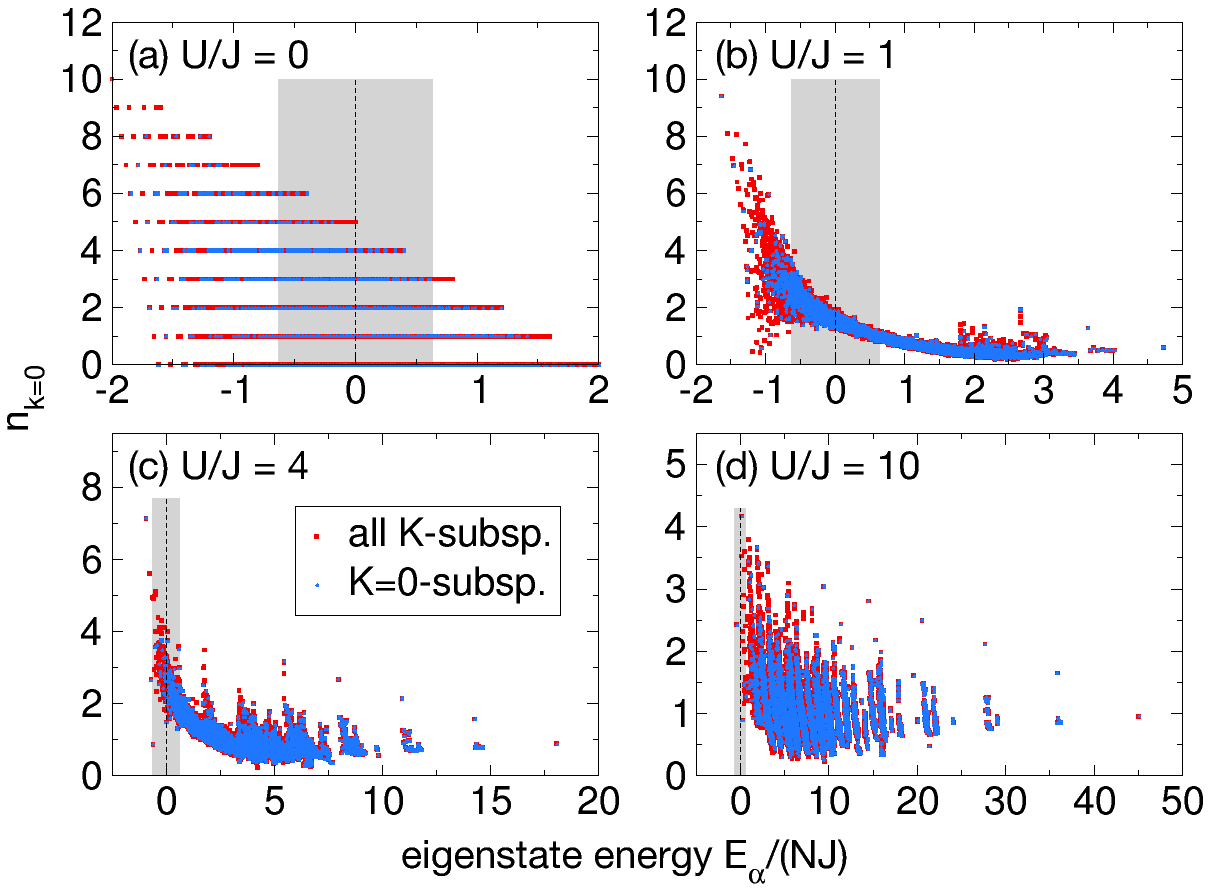}
        \caption{(Color online) Eigenstate expectation values as in Fig. \ref{fig:mel1}, but for the MDF at  $k=0$, $\bra{\alpha}\hat{n}_{k=0}\ket{\alpha}$. Note that in the integrable case $U/J=0$, the distribution of energy eigenstate expectation values of the global observable $n_{k=0}$ is very broad. Results for $\bra{\alpha}\hat{n}_{k=0}\ket{\alpha}$  were shown and discussed in Ref.~\cite{roux10}.}
        \label{fig:mel2}
\end{figure}

\begin{figure}[t]
        \includegraphics[width=0.49\textwidth]{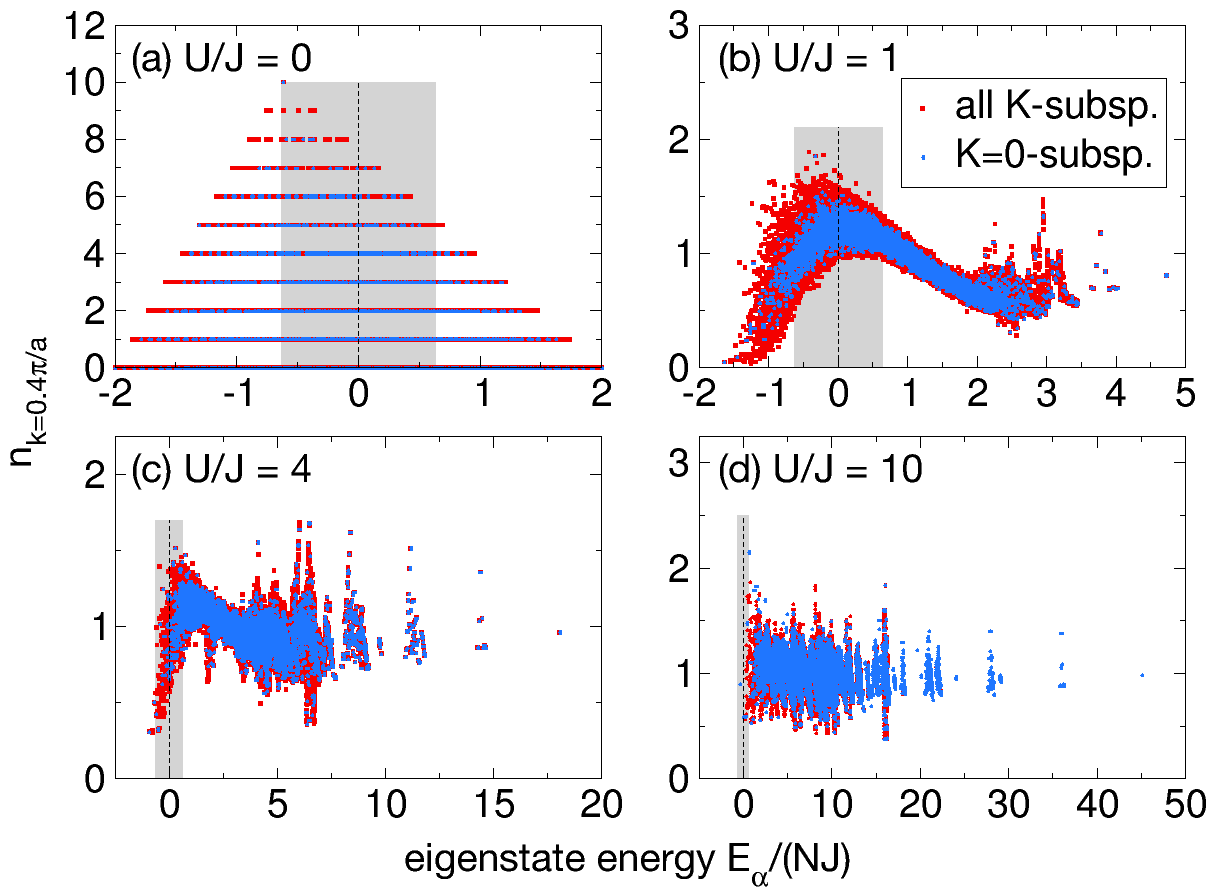}
        \caption{(Color online) Eigenstate expectation values as in Fig. \ref{fig:mel1}, but for the MDF at  $k a=0.4\pi$, $\bra{\alpha}\hat{n}_{k=0.4\pi/a}\ket{\alpha}$.}
        \label{fig:mel3}
\end{figure}

\begin{figure}[htbp]
        \includegraphics[width=\columnwidth]{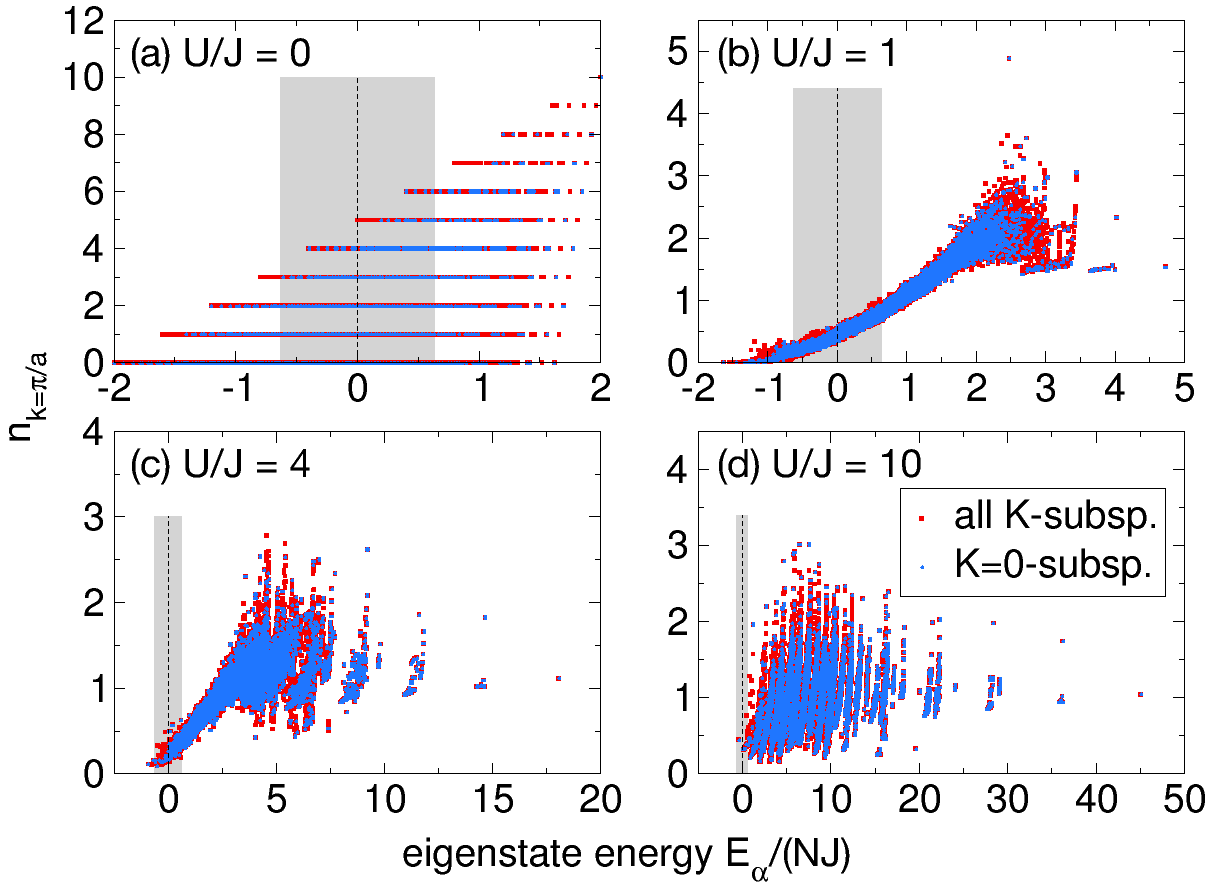}
        \caption{(Color online) Eigenstate expectation values as in Fig. \ref{fig:mel1}, but for the MDF at  $ka =\pi$, $\bra{\alpha}\hat{n}_{k=\pi/a}\ket{\alpha}$.
        }
        \label{fig:mel4}
\end{figure}

To that end, we compute the expectation values
of $\nu_h$ and $n_k$ in all eigenstates $\ket{\alpha}$ of the post-quench Hamiltonian. 
Our results are shown in Fig.~\ref{fig:mel1} and 
Figs.~\ref{fig:mel2},~\ref{fig:mel3} and~\ref{fig:mel4}  for $U/J=0,1,4,10$.
For the MDF, we present results for three different quasi-momenta $k=0, 0.4\pi/a, \pi/a$ in Figs.~\ref{fig:mel2},~\ref{fig:mel3} and~\ref{fig:mel4}, respectively. 
In these four figures we show both the full distributions and the distribution restricted to the total quasi-momentum $K=0$ subspace the initial state lies in.

Several observations need to be stressed. First, for $U=0$ and $U=J$, the system's energy $E=0$ is in the center of the spectrum, while
for large $U/J$, it sits at the lower edge.
Hence, finite-size effects are expected to be more severe at large $U/J$~\cite{roux10}
(which corresponds to the small-quench regime discussed in \cite{roux09})
and therefore, one would expect 
thermalization to be observable numerically---if at all---only on comparably large systems.
Second, both for $U=0$ and large $U/J$, the distributions of $\langle \alpha| \hat n_k|\alpha \rangle $ are very broad and structured.
This is typical for integrable systems or systems that are close to an integrable point \cite{rigol08,rigol09}. 
At $U=0$, this is a consequence of integrability and the existence of many degeneracies. For $n_k$, the particular
structure at $U=0$ emerges by choosing the simultaneous eigenbasis of all $n_k$, in which the $\langle \alpha| \hat n_k|\alpha \rangle $
take only integer numbers.
For large $U/J$, 
 the structure of the spectrum at the integrable $U/J=\infty$ point (compare \cite{kollath10,roux10})
is reflected in the distributions of $\langle \alpha| \hat \nu_h|\alpha \rangle $, with separate bands emerging, corresponding to 
excitations with an energy $n U$ above the ground state ($n$ integer). 
Third, for $U=0$, $\langle \alpha|\hat \nu_h| \alpha \rangle $ exhibits almost no dependence on energy.

An important requirement for the ETH is that $\langle \alpha | \hat A | \alpha \rangle $ is a smooth and sharp function of energy $E$.
This means that the {\it vertical} width of the distribution of  $\langle \alpha | \hat A | \alpha \rangle $ at a fixed $E$ should be small, compared to the 
typical size and overall variation of  $\langle \alpha | \hat A | \alpha \rangle $ in that region.
By inspection of Figs.~\ref{fig:mel1},~\ref{fig:mel2},~\ref{fig:mel3} and~\ref{fig:mel4}, this condition seems to be mostly fulfilled  for $U=J$ with the exception of $n_{k}$ at $k=0.4\pi/a$ (see Fig.~\ref{fig:mel3}(b)). For $U/J=4$ and $U/J=10$, the typical vertical width of the distribution across the entire spectrum
is much larger, yet  at $E=0$, the eigenstate expectation values have smooth distributions compared to the behavior at larger $E$. 

For $U=0$, there are always several significantly different expectation values for a given value of $E_{\alpha}$ (except for $\nu_h$), and thus one
would in general not expect thermal behavior for $n_k$.
At present, we do not understand why the distribution of $\langle \alpha | \hat \nu_h | \alpha \rangle $ has such a
simple structure at $U=0$, namely a narrow distribution with $\langle \alpha | \hat \nu_h | \alpha \rangle$ almost independent of energy
(compare Fig.~\ref{fig:mel1}(a)). 
Qualitatively, since the Hamiltonian at $U=0$ does not depend on $\hat n_i$ or any higher occupancies,
it implies that the eigenenergies are also independent of higher occupancies and thus $\nu_h$ is only a 
function of average density but not of eigenstate energy. Nonetheless there are several flat bands of eigenstates, all of which have different
values of $\nu_h$, with no dependence on $E_{\alpha}$.

\begin{figure}[t]
        \centering
        \includegraphics[width=\columnwidth]{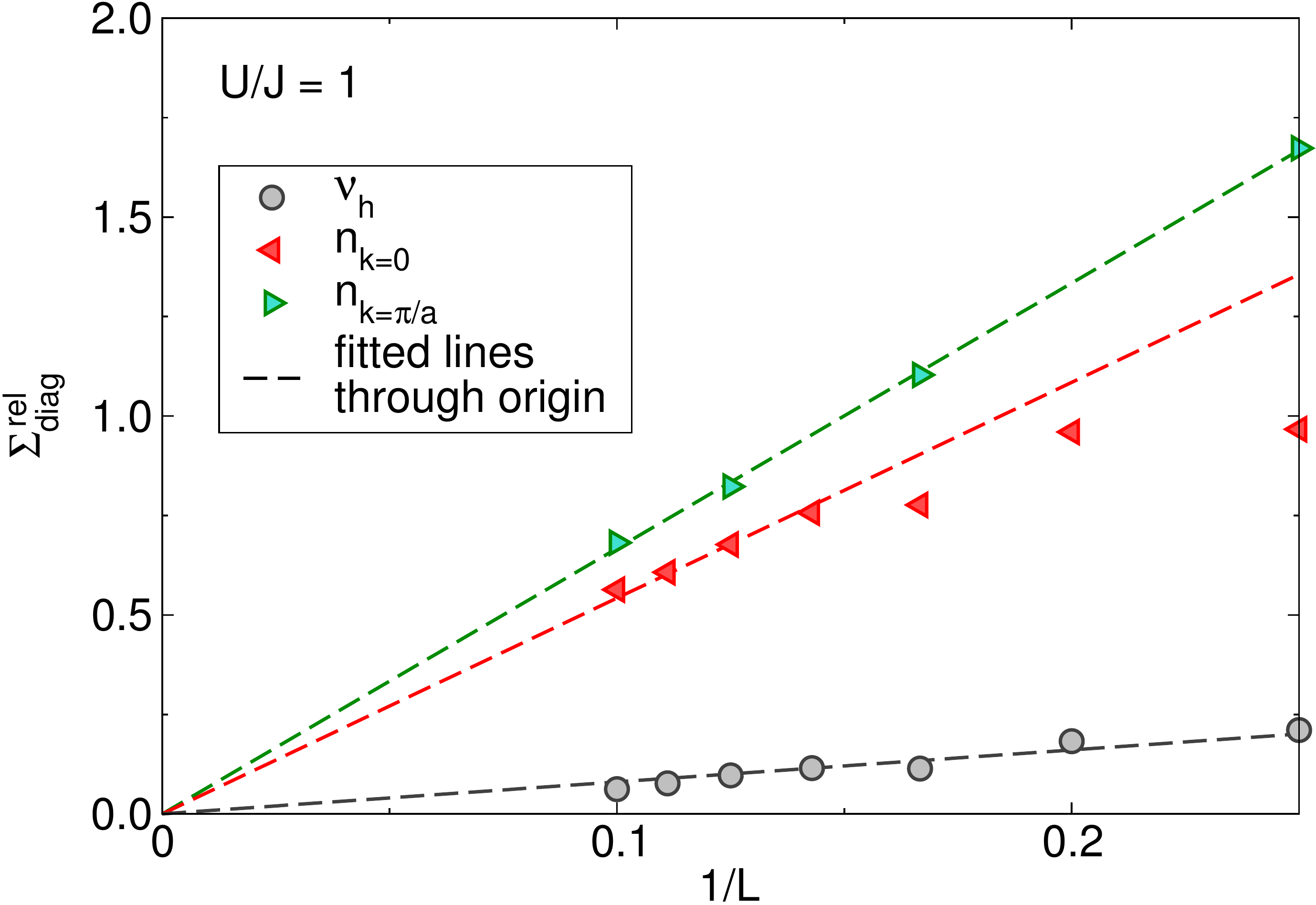}
        \caption{ (Color online)
        $\Sigma^\text{rel}_\text{diag} = \Sigma_{\delta A, \text{diag}} / \braket{\hat A}_{\rm diag}$ as a function of inverse system size $1/L$ for
         different observables at $U/J=1$. Dashed: Linear fits  through origin (for $n_{k=0}$, the three smallest system sizes are excluded from the fit).}
        \label{fig:ETH_validity}
\end{figure}

To understand our results obtained for a fixed system of finite size and for our specific initial state, we are interested in the deviation of the eigenstate expectation values with respect to the diagonal ensemble,
$\delta A_{\alpha} = A_{\alpha\alpha} - \braket{\hat{A}}_{\rm diag}$.
We calculate the variance of $\delta A_{\alpha}$ in the diagonal ensemble as
\begin{equation}
\Sigma^2_{\delta A,\rm diag} = \sum_\alpha |c_\alpha|^2 \left(\hat A_{\alpha\alpha}\right)^2 - \left(\sum_\alpha |c_\alpha|^2 \hat A_{\alpha\alpha}\right)^2.
\end{equation}
Our results are shown in Fig.~\ref{fig:ETH_validity}, where we actually plot
\begin{equation}
\Sigma^{\rm rel}_{\rm diag}= \Sigma_{\delta A, \rm diag}/\langle \hat A\rangle_{\rm diag}\,.
\end{equation}
We observe, similar to related studies \cite{rigol08,steinigeweg13,beugeling14,steinigeweg14}, that these fluctuations
decay quickly with system size in this non-integrable model. However, note that $\Sigma^\text{rel}_\text{diag} \sim 1/L$
(i.e.,  power-law dependence on $1/L$), in contrast to other measurements of the vertical width of 
distributions of $\langle \alpha | \hat A | \alpha \rangle$ studied in the literature \cite{beugeling14}.
There, one usually desires an initial-state independent criterion and therefore, one studies the fluctuations of the eigenstate expectation values relative to the corresponding micro-canonical values, $\Delta A_\alpha = A_{\alpha\alpha} - \braket{\hat{A}}_{\rm mc}(E_\alpha)$.
The variance of these fluctuations
\begin{equation}
\Sigma^2_{\Delta A} = \langle (\Delta A_\alpha)^2\rangle_{\rm c} -  \langle \Delta A_\alpha \rangle_{\rm c}^2\,
\end{equation}
is typically measured with respect to the central region of the spectrum $\langle ... \rangle_{\rm c}$, which contains the most eigenvalues.
This quantity decays as $\Sigma_{\Delta A} \sim 1/\mathcal{D}^{1/2}$ \cite{beugeling14}, where $\mathcal{D}$ is the Hilbert-space dimension,
and thus exponentially fast in $L$.

The difference clearly is that our measure $ \Sigma_{\delta A,\rm diag}$ also takes into account the width
of the initial state, which, as our examples show, can contribute to deviations from thermal behavior
on small systems, as was also emphasized in \cite{biroli10}. We believe that it is important to consider this aspect in the analysis
of numerical data at finite $L$ for specific examples of quantum quenches, while in order to assess the validity of
 the ETH assuming sufficiently narrow initial states (which will usually be the case \cite{rigol08}), $\Sigma_{\Delta A}$ provides the  appropriate measure.

\subsection{Micro-canonical ensemble}
\label{sec:mc}
The results from the previous section Sec.~\ref{sec:eth}
on the ETH suggest that the micro-canonical ensemble should yield a reasonable description for
most observables in the $0<U/J \lesssim 5$ regime, where most observables have (i) narrow distributions of eigenstate
expectation values $\langle \alpha | \hat A| \alpha   \rangle $ and (ii) the total energy $E$ is in the bulk 
of the system. Of course, neither the fluctuations of the eigenstate expectation values are vanishingly small  for $L=10$ nor
is the initial state narrow in energy. In fact, for $L=10$, $\sigma_{\rm diag}\approx 0.63 JN$. 

The results for $U=0$ and $U/J=1$ shown in Figs.~\ref{fig:mel1}(a) and~\ref{fig:mel1}(b), respectively, suggest that $(\nu_h)_{\alpha\alpha}$ can be reasonably approximated as a linear function of energy.
As a direct consequence of this observation, the micro-canonical average matches the diagonal one even though $L=10$ is small.
A similar behavior, i.e., a linear variation with $E_\alpha$ in the vicinity of $E=0$,  is realized 
for the MDF for certain values of the quasi-momentum $k$ (see,  e.g., Fig.~\ref{fig:mel4}(b)).

Nonetheless, there are cases in which the state as such is broad in energy such that 
a large variation of $\langle \alpha | \hat A| \alpha   \rangle $ with energy is resolved (see Fig.~\ref{fig:mel3}(b)),
resulting in 
$$
\langle \hat A \rangle_{\rm diag } \not= \langle \hat A \rangle_{\rm mc }\,. 
$$
This is clearly a finite-size effect since ultimately $\sigma_{\rm diag}/(NJ) \to 0$ (note that we plot $\langle \alpha | \hat A| \alpha   \rangle $
versus $E_\alpha/N$  in Figs.~\ref{fig:mel1}-\ref{fig:mel4}).

The failure of the micro-canonical ensemble at large $U/J$ is due to the fact that
the total energy lies in the gap $\Delta$ for these small system sizes. Therefore, in the vicinity
of $E=0$ there are no states at all, and a naive implementation of the micro-canonical ensemble
will mix in the ground state and states from the first band, which contains states with one doublon
excitation. As a consequence, one can easily convince oneself that
$\langle \hat \nu_h \rangle_{\rm mc} > \nu_h$.
This, of course, is a finite-size effect,
since the quench energy in a global quench is extensive \cite{roux10}, while the Mott gap $\Delta $ is intensive.

Nevertheless, the system's energy will sit in the bulk of the spectrum if the limit  $N,L\to \infty$ at $n=N/L=1$ is taken at $U/J<\infty$.  
In this case, one would expect that the comparison to thermal ensembles becomes meaningful again.
We quantify this condition as $\delta E \gg \Delta$, following  Ref.~\cite{roux10}.
To estimate the number of particles $\tilde N$ which satisfy this condition, we use the following approximations:
(i) We obtain the quench energy from  second-order perturbation theory.
For periodic boundary conditions,  this results in  $\delta E = \frac{4J^2}{U} N$.
(ii) We use the effective holon-doublon quadratic model~\cite{cheneau12,queisser14} to estimate the excitation gap as $\Delta =  U - 6J$.
The requirement $\delta E \gg \Delta$ then leads to
\begin{equation} \label{ncond}
\tilde N \gg \frac{1}{4} \frac{U}{J} \left(\frac{U}{J} -6\right).
\end{equation}
For $U/J=10$, this implies $\tilde N \gg 10$, while for $U/J=20$, it implies $\tilde N \gg 70$.
Alternatively, one could require that the width of the diagonal distribution $\sigma_{\rm diag}$, Eq.~(\ref{eq:sigma_width}), is much larger than the gap $\Delta$.
This leads to
\begin{eqnarray}
\tilde L & \gg & \frac{1}{4} \left( \frac{U}{J} - 6\right)^2. \label{lcond}
\end{eqnarray}
For large $U/J$ and $N=L$, the latter condition becomes very similar to the condition in Eq.~(\ref{ncond}).
Evidently, in a typical quantum gas experiment, in which there are about $N\sim 100$ particles per tube, thermalization 
in the small quench regime can thus never occur.
 
\begin{figure}[!tbp]
        \centering
        \includegraphics[width=\columnwidth]{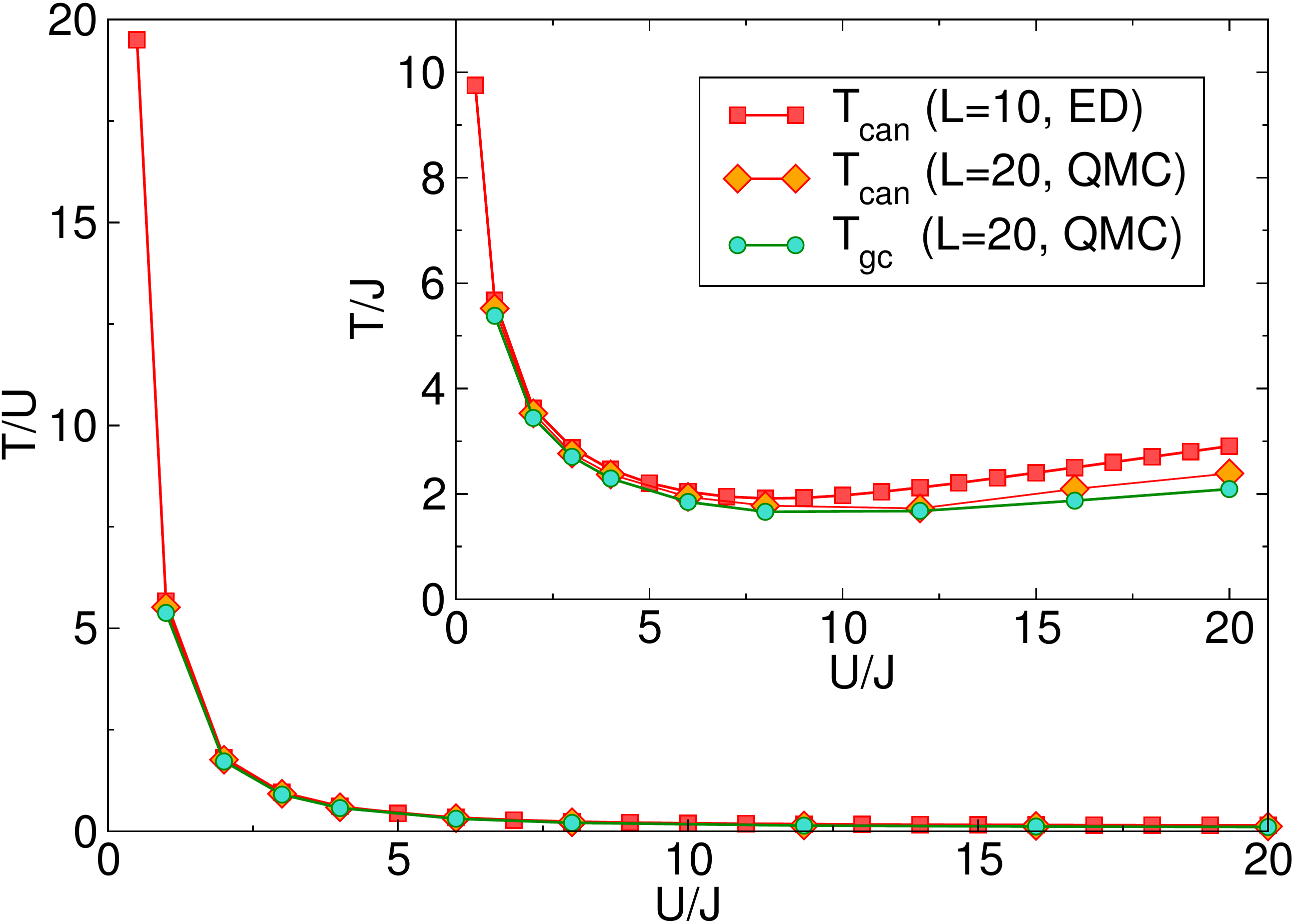}
        \caption{
        (Color online) Temperatures of the canonical and the grand-canonical ensembles as a function of interaction strength $U/J$ for  systems with energy $E = 0$ and PBC. Main panel: $T/U$ versus $U/J$, inset: $T/J$ versus $U/J$. The data for $L=10$ was obtained with exact diagonalization (ED), the ones for $L=20$ are Quantum Monte Carlo (QMC) results.
       } \label{fig:Teff}
\end{figure}

\subsection{Canonical ensemble}
\label{sec:can}

\subsubsection{Temperature}
In the canonical ensemble, it is always possible to 
enforce $E = \langle H \rangle_{\rm can} $ by choosing temperature.
Whether the temperatures extracted from this condition are meaningful for a small
system is a different question though. 
One may require several criteria such as (i) independence of $T_{\rm can}$ on system size
and, even stronger, (ii) equivalence of  definitions of $T$ in different ensembles \cite{roux10a}.

Figure~\ref{fig:Teff} shows the canonical temperature $T_{\rm can}$ for two system sizes $L=10$ and $20$
and it also includes the grand-canonical temperature $T_{\rm gc}$ for $L=20$ (note that our results differ qualitatively from Ref.~\cite{barmettler13}).
In the main panel where we plot $T/U$ versus $U/J$, we simply observe that $T$ increases monotonically
from $T=0$ at $U/J=\infty$ towards $U=0$, where the temperature diverges. This behavior  follows the 
dependence of the excess energy $\delta E$.

The inset, which displays $T/J$ versus $U/J$, unveils the finite-size effects: These are negligible 
for $U/J \lesssim 5$ but $T_{\rm can}(L=10)> T_{\rm can}(L=20)$ as $U/J$ increases. At the same time,
$T_{\rm gc}$ is practically identical to  $T_{\rm can}(L=20)$  for $U/J \lesssim 5$, 
but again deviates from that for large interaction strengths. 
On the one hand, it is very interesting that at small $U/J \lesssim 5$ and system sizes of $L \sim 20$
there is already a notion of ensemble equivalence with $\langle \hat A \rangle_{\rm mc }  \approx \langle \hat A \rangle_{\rm can } $
for at least some observables (e.g., $\nu_h$) and agreement between different definitions of temperature.
On the other hand, these results suggest that the similarity between canonical and diagonal ensemble
at large $U/J$ has to be taken with some caution since (i) the micro-canonical ensemble cannot properly be constructed and thus there is no ensemble equivalence and (ii) canonical temperatures are evidently system-size dependent.

Since the canonical temperature at $U=0$ is $T=\infty$, one can easily understand the apparent
agreement between the canonical and diagonal expectation values of $\hat n_k$ as a coincidence.
Namely, as a consequence of the initial state being a product state plus the fact that the $n_k$ are integrals of motion for $U/J=0$, $n_k(t=0)=n_k(t)=\braket{\hat n_k}_\text{diag}$, $n_k$ immediately has the right $T=\infty$ form, i.e., $n_k=N/L$.

\subsubsection{Statistical weights of the initial state}

We now turn to an analysis of the weights of the diagonal and canonical ensemble and the respective energy distributions.
This will be instrumental in  understanding why the canonical ensemble works so well for $U/J \lesssim 5$, and we thus start our
discussion with the small $U/J$ regime.

\begin{figure}[t]
        \centering
        \includegraphics[width=.96\columnwidth]{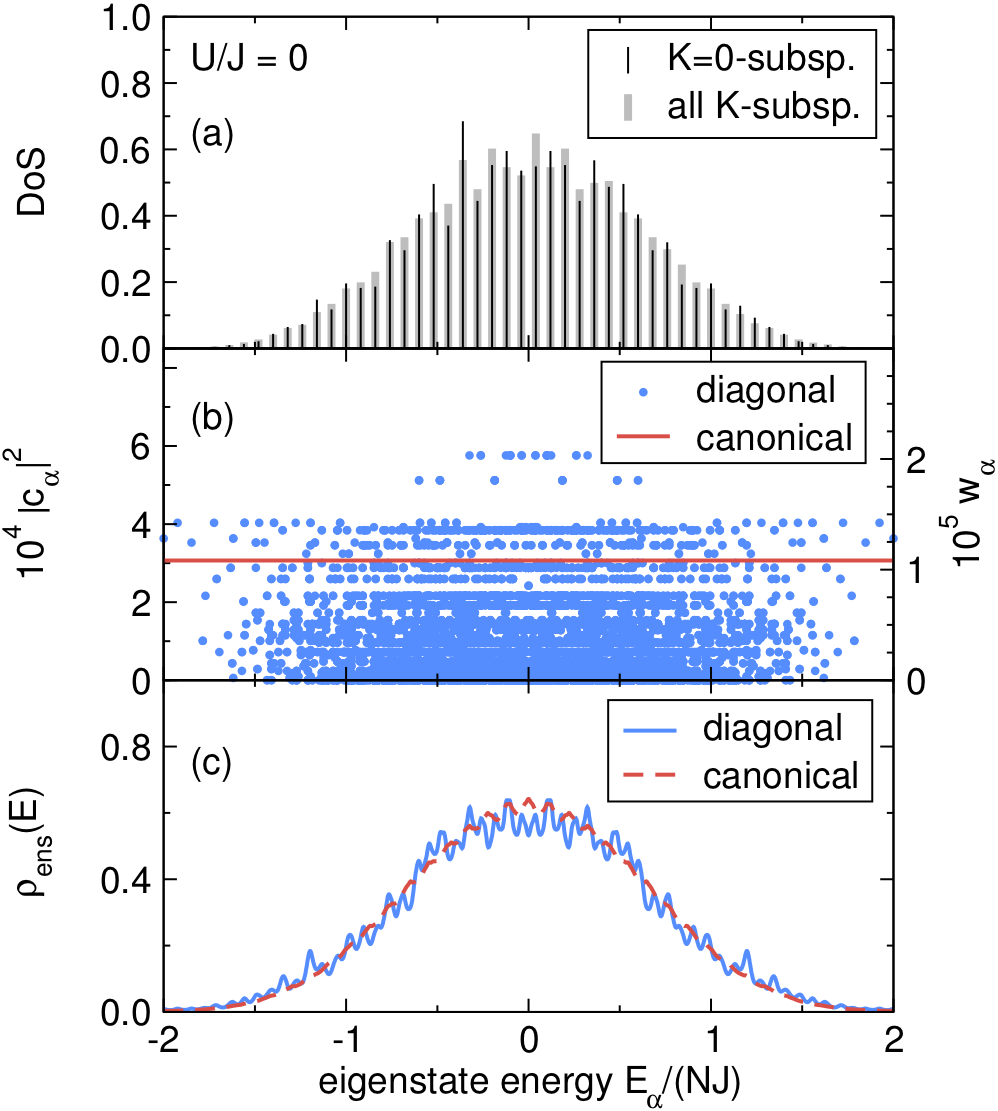}
        \caption{(Color online) Comparison of the energy distributions in the diagonal and the canonical ensembles for the quench $U/J=\infty \rightarrow U/J=0$ ($L=10$, PBC). {(a)} Density of states (DoS) as a histogram (arb. units) of the number of eigenstates in 50 bins for the $K$=0 subspace, which is the only one contributing to the diagonal ensemble for our particular quench initial state, and for all $K$-subspaces, since all states contribute to the canonical ensemble. The histogram for $K$=0 is multiplied by $L$ to simplify the comparison with the DoS of all $K$-subspaces. The DoS in the $K$=0 subspace is very similar to the one for all $K$-subspaces. {(b)} Distribution of weights in the diagonal  ($|c_\alpha|^2$)  and in the canonical ensemble ($w_{\alpha} = e^{-\beta E_{\alpha}}/Z$). 
The canonical distribution is flat  because $T_\text{can}=\infty$. {(c)} Energy distribution $\rho_{\rm ens}(E)$ (arb. units), which is the product of the eigenstate occupations and the density of states. The $\delta$-peaks at each $E_\alpha$ are broadened by Lorentzians $1/\left( \pi\gamma \left(1 + \left(\frac{E-E_\alpha}{\gamma}\right)^2\right) \right)$ of width $\gamma = 0.02 NJ$. $\rho_\text{diag}(E)\approx \rho_\text{can}(E)$ since both the DoS and the eigenstate occupations are similar for both ensembles.
      The DoS in the 1D BHM for various $U/J$ was previously studied in Ref.~\cite{kollath10}.}\label{fig:DoSU0}
\end{figure}

\begin{figure}[t]
        \centering
        \includegraphics[width=.96\columnwidth]{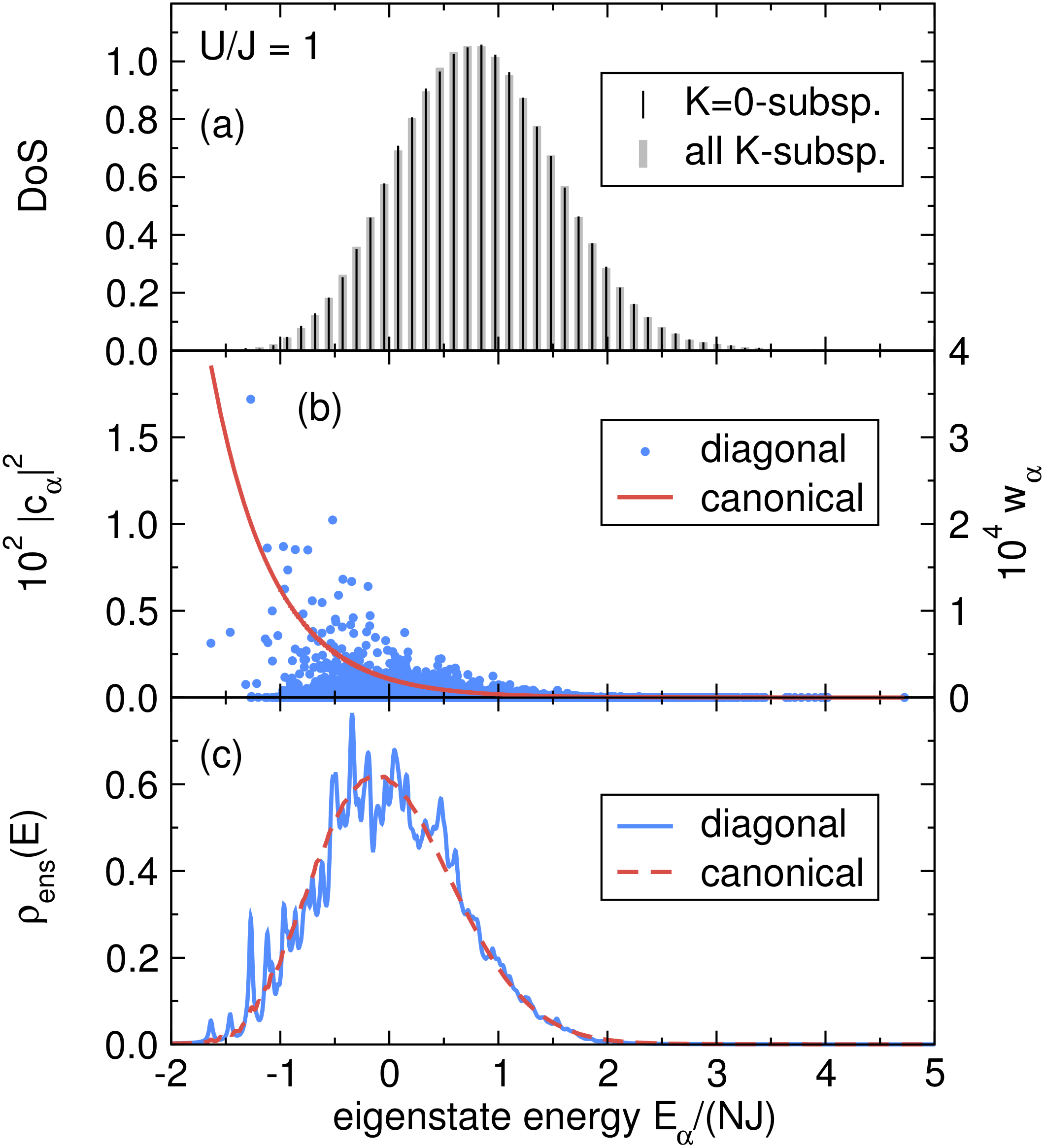}
        \caption{(Color online) Comparison of the energy distributions in the diagonal and the canonical ensembles as in Fig. \ref{fig:DoSU0}, but for the quench $U/J=\infty \rightarrow U/J=1$. {(a)} The DoS in the $K$=0 subspace is essentially equal to the one for all $K$-subspaces. {(b)} The eigenstate occupations of the two ensembles obviously  differ, with the exponential decay in the canonical ensemble being clearly visible. {(c)} $\rho_\text{diag}(E)\approx \rho_\text{can}(E)$ is in this case due to the interplay of DoS and eigenstate occupations.
        }
        \label{fig:DoSU1}
\end{figure}

{\it Small U/J.}
We present typical examples for the energy distribution in Figs.~\ref{fig:DoSU0} and \ref{fig:DoSU1} for $U=0$ and $U/J=1$, respectively. Evidently, for small $U/J$, this distribution is centered around
$E=0$ in the sense that it has its maximum in the vicinity of $E=0$ 
and  we observe a Gaussian shape of $\rho_{\rm diag}(E)$. We will  complement this observation by the discussion of the
short-time dependence of the fidelity in Appendix~\ref{sec:fidel}.

 We also include the canonical energy distribution        in Figs.~\ref{fig:DoSU0} and \ref{fig:DoSU1}, defined in  Eq.~\eqref{eq:can_en_distr}. 
Curiously, we find that
\begin{equation}
\rho_{\rm can}(E) \approx \rho_{\rm diag} (E)\,.
\label{eq:candiag}
\end{equation}
From this observation, we also expect
\begin{equation}
\langle \hat A \rangle_{\rm can}= \langle \hat A \rangle_{\rm diag}\,
\label{eq:dc2}
 \end{equation}
for any observable $\hat A$ and the {\it special} initial state considered here. 
A similar observation for quench dynamics with 1D hard-core bosons  starting from a {\it product } state in real space
was discussed in Ref.~\cite{rigol11}.

\begin{figure}[!t]
        \centering
        \includegraphics[width=.8\columnwidth]{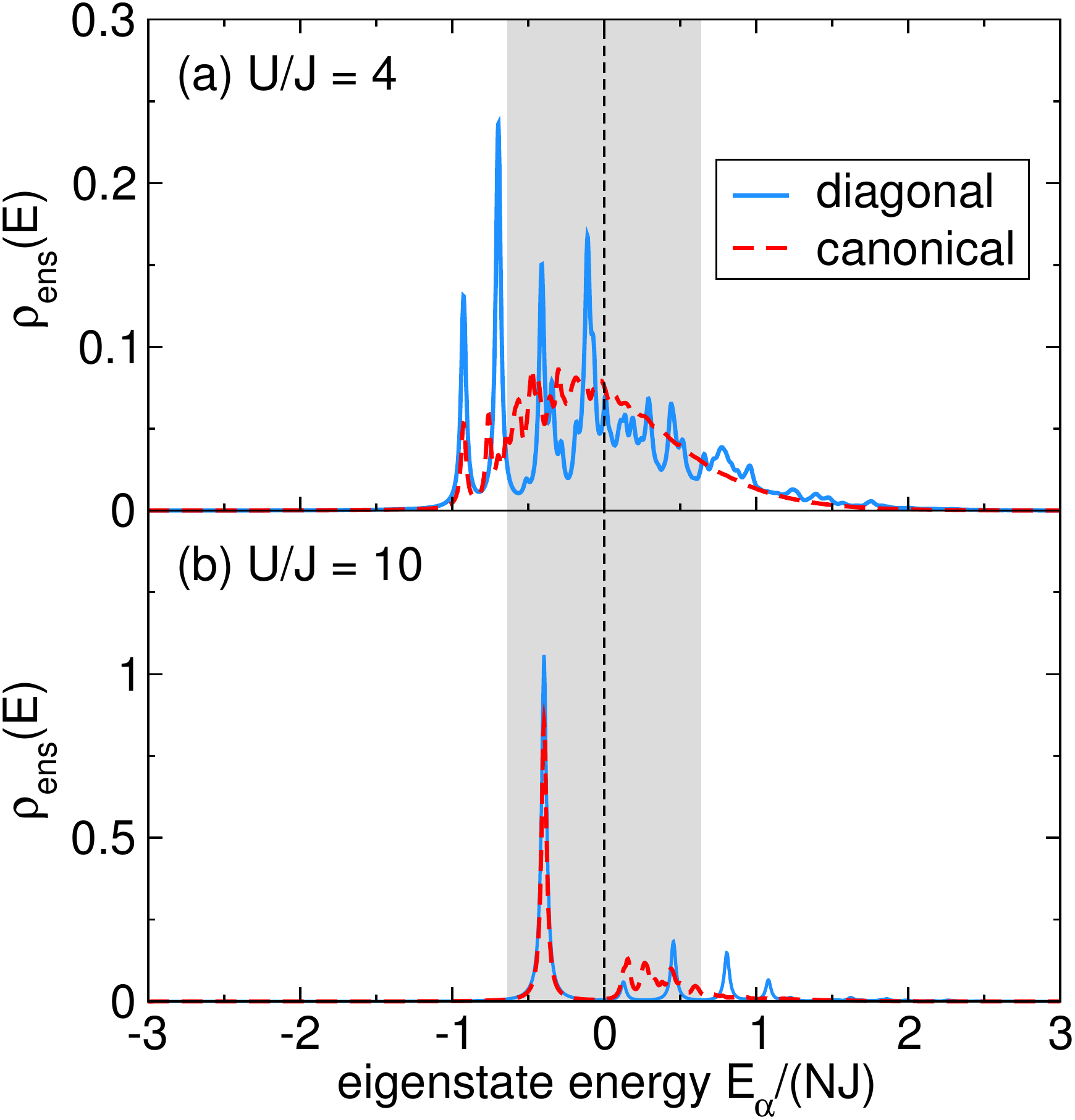}
        \caption{(Color online) Energy distributions (arb. units) in the diagonal and the canonical ensembles as in Fig.~\ref{fig:DoSU0}(c) and \ref{fig:DoSU1}(c), but for the quenches to (a) $U/J=4$ and (b) $U/J=10$. For $U/J=10$, the system's energy (vertical dashed lines) lies in the gap between the ground and the first excited state, rendering  the micro-canonical ensemble inapplicable on such small systems. Shaded area: Width of the initial state.
        }
        \label{fig:rho_U4_U10}
\end{figure}

As the simplest case to understand this observation, we first consider $U/J=0$. 
Since $T_{\rm can}=\infty$ for our initial state, the canonical density matrix is equal to
\begin{equation}
\hat \rho_\text{can} = \frac{1}{Z} \sum_{\alpha} e^{-\beta E_\alpha} \ket{\alpha}\bra{\alpha} = \frac{1}{\cal D} \sum_{\alpha} \ket{\alpha}\bra{\alpha}\,
\end{equation}
where $\cal D$ denotes the dimension of the Hilbert space.
The width of the canonical distribution is therefore directly given by the width of the density of states (DoS) $g(E)$.
In Fig.~\ref{fig:DoSU0}, we show $\rho_{\rm can}(E)$, 
which can faithfully be fitted by a Gaussian function with $\sigma_{\rm can} \approx 0.63 JN$.
This leads to 
\begin{equation}
\sigma_{\rm can} \approx \sigma_{\rm diag}
\end{equation}
for our initial state and $L=10$ and the remarkable agreement of Eq.~\eqref{eq:candiag}.
Even though $\hat \rho_{\rm can}$ and $\hat \rho_{\rm diag}$ differ on the level of weights of single eigenstates (see Fig.~\ref{fig:DoSU0}(b)), the ensemble averages result in virtually indistinguishable expectation values.
This is an example of the case, in which the coefficients $c_{\alpha}$ fully sample the energy shell (reflected in the Gaussian
energy distribution) \cite{torres-herrera14,torres-herrera13},
which also leads to initial-state independence in Eq.~\eqref{eq:gibbs}.
According to Ref.~\cite{torres-herrera13}, this is sufficient to ensure that steady-state values of observables  agree
with thermal ones at infinite temperature.
%Also note that studies of the level-spacing distriution conclude that the spectrum of the 1D Bose-Hubbard model is chaotic for small $U/J$ \cite{kollath10,buchleitner}.

The corresponding temperatures $T_{\rm can}$ decrease as a function of $U/J$, as shown in Fig.~\ref{fig:Teff} for $L=10$ and $L=20$.
Assuming that the density of states $g(E)$ is a Gaussian function centered at the energy $E_{\rm DoS}$, one can show that the canonical distribution $\rho_{\rm can}(E)$ is, for any $T_{\rm can}$ chosen to match a given $E$, also a Gaussian function (of the same width), yet 
 centered at  $E$.
We therefore expect that the origin of the agreement between the canonical and diagonal averages is the same as in the $U/J=0$ case described above, i.e., similar widths (and similar higher moments) of $\rho_{\rm can}(E)$ and $\rho_{\rm diag}(E)$ result in virtually indistinguishable ensemble averages.
This can be viewed as being related to  the emergence of quantum chaos in the one-dimensional Bose-Hubbard model in the small $U/J$ regime \cite{kollath10,buchleitner}.

In fact, one can expect similar properties for
any initial state that is a product state in real space: 
The energy at $U=0$  vanishes for all these states, hence the canonical temperature is always infinite.
Moreover, the eigenstates at $U=0$ are built from single-particle eigenstates of the $\hat n_k$ and it is therefore not suprising that
 these are sampled practically randomly by initial states that are real-space product states.

{\it Large U/J.}
At large $U/J>4$, Eq.~\eqref{eq:candiag} is not valid and neither the DoS nor the energy distribution 
is Gaussian. Rather,   the distribution $\rho_{\rm diag}(E)$
has a two-peak structure with most of the weight sitting in the ground state and in the states with $E_{\alpha }> E=0$, while there are virtually no
states in the vicinity of $E=0$ (see Fig.~\ref{fig:rho_U4_U10}(b)). This is a consequence of $E=0$ being in the gap deep in the Mott-insulating regime on finite systems, where as a consequence, the {density of states}
at  $E=0$ vanishes.
The fact that $\nu_h$ is well described by the canonical ensemble actually does not require the energy distributions  to be similar,
only the overall weight in each band with a fixed number of excitations above the ground state has to be identical (see Fig.~\ref{fig:rho_U4_U10}(b) and Sec. \ref{subsec:fraction}).
Since at large $U/J$ and on the accessible system sizes, ETH does not apply, the apparent similarity of the diagonal and canonical
ensemble has a different origin, not covered by ETH, namely the peculiar structure of $\hat \rho_\text{diag}$.

\begin{figure}[t]
        \centering
        \includegraphics[width=.96\columnwidth]{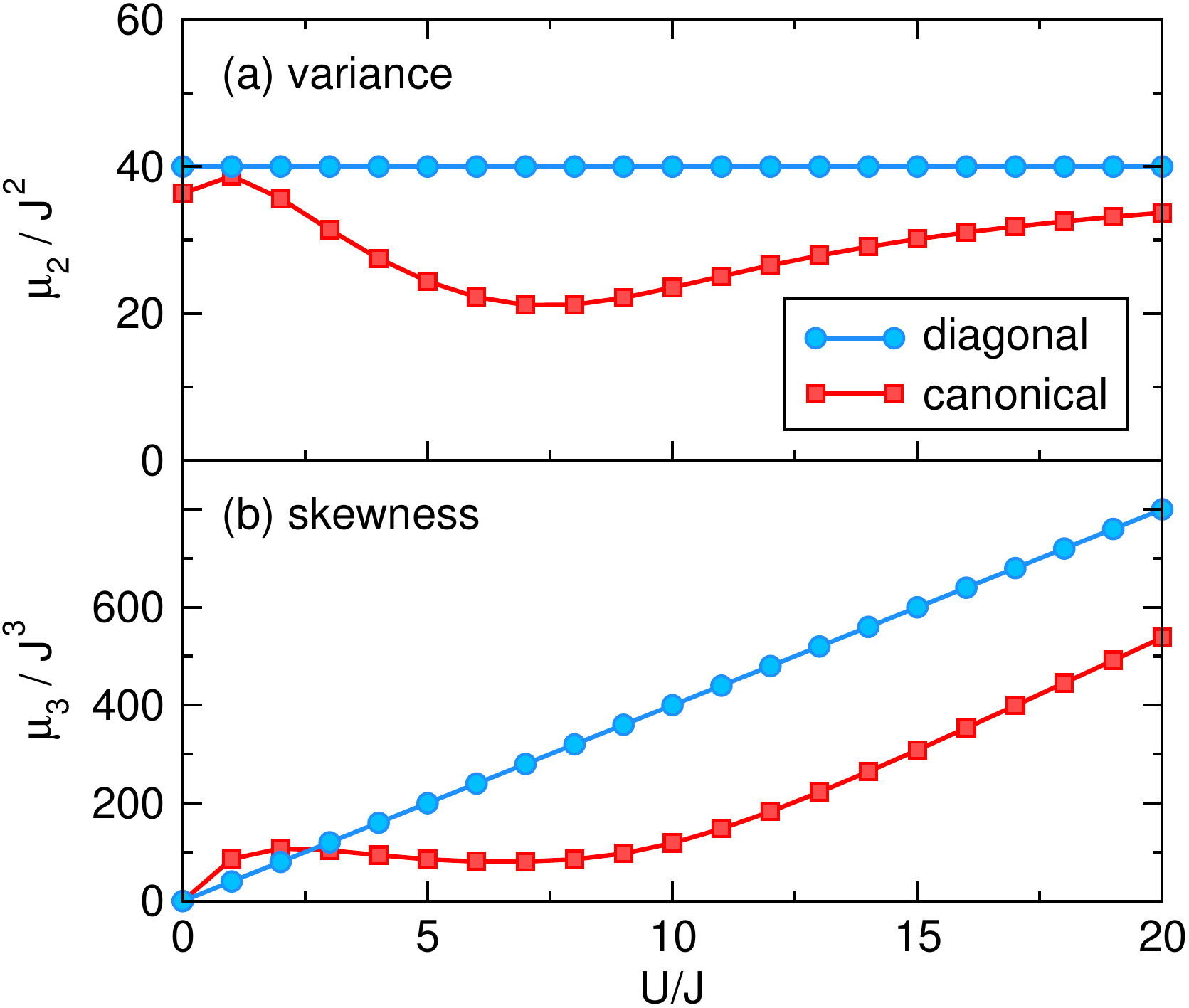}
        \caption{
        (Color online) Central moments $\mu_2$ (variance) and $\mu_3$ (skewness) (see Eq.~\eqref{eq:moments}) of the energy distribution for the diagonal and the canonical ensemble (ED data, $L=10$, PBC).}
        \label{fig:moments}
\end{figure}

In order to get a quantitative handle on how the agreement between $\rho_{\rm diag}(E)$ and $\rho_{\rm can}(E)$
evolves as $U/J$ is increased, we calculate the central moments
\begin{equation}
\mu_n = \braket{(E-\braket{E}_{\rm ens})^n}_{\rm ens}
\label{eq:moments}
\end{equation}
for both distributions (ens = diag, can).
For a normal distribution $\mu_2\not= 0$ and $\mu_3=0$, while for the diagonal distribution $\mu_2/J^2 = 4L$ and $\mu_3/J^3=4L(U/J)$ by straightforward calculations.
From Fig.~\ref{fig:moments}, which shows the variance
$\mu_2$ and the skewness $\mu_3$ versus $U/J$, we infer
that (i) the larger $U/J$, the less the diagonal and canonical energy distributions resemble a normal distribution,
(ii) the distributions become skewed, indicated by non-zero odd moments, and (iii) the second moments of the diagonal and canonical energy distributions
are similar at small and large $U/J$.
The last point is consistent with our observation that the relative difference between diagonal and canonical ensemble
is the largest at intermediate $U/J$ (compare Fig.~\ref{fig:tav}),
namely of the order of 10\%.

\subsubsection{Comparison between diagonal and canonical expectation values and finite-size scaling}

The results for $L=10$ and small $U/J <4$ do not imply a general equivalence of expectation values since $\sigma_{\rm can}$ and $\sigma_{\rm diag}$ do not necessarily coincide for all values of $L$. To further elucidate the significance of the similarity between the diagonal and
the canonical ensemble, we study the finite-size scaling in more detail.

\begin{figure}[!t]
        \centering
        \includegraphics[width=0.96\columnwidth]{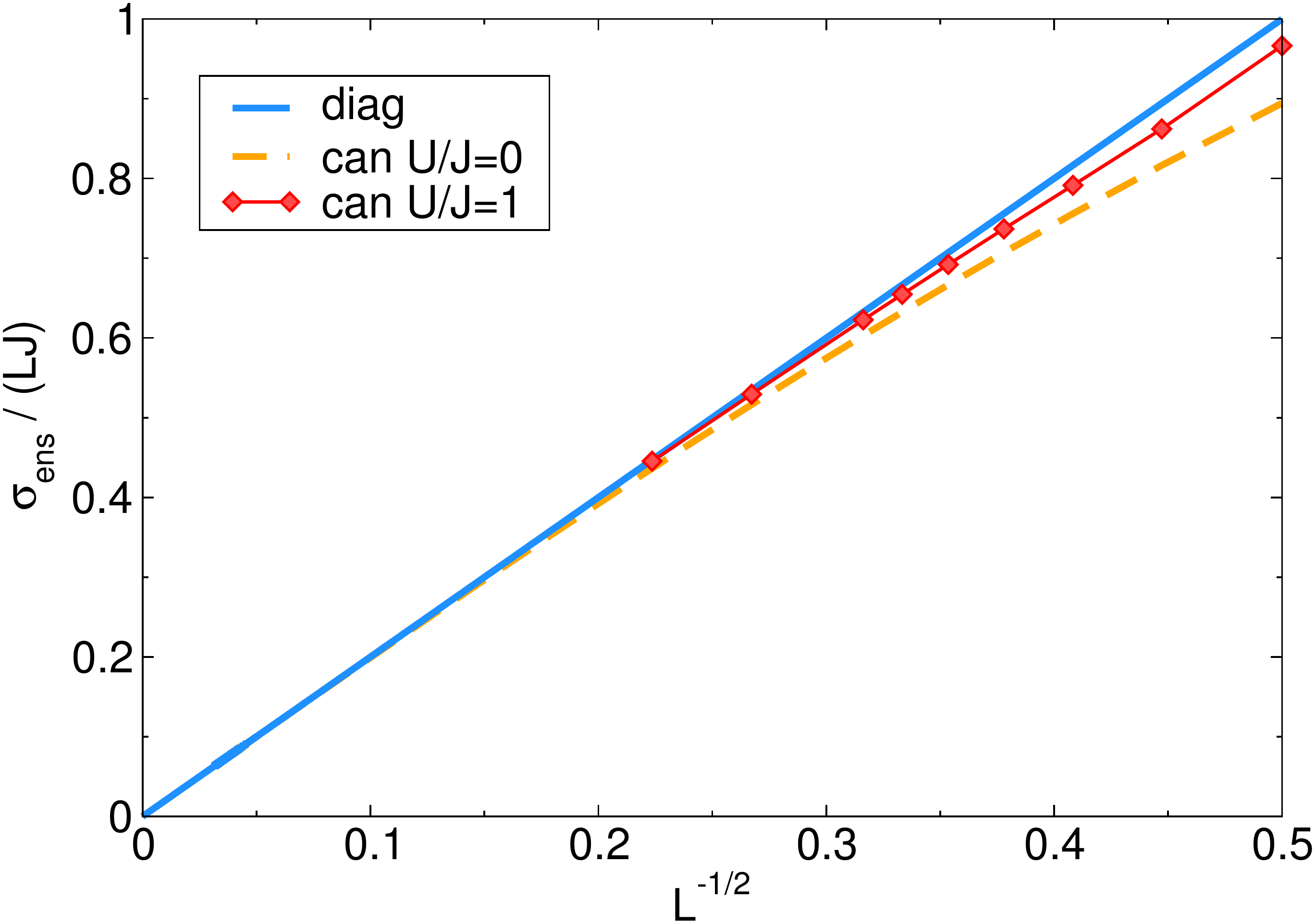}
        \caption{(Color online) Width $\sigma_{\rm ens}$ of the energy distributions  (ens = diag, can) as a function of system size $L$.
We plot $\sigma_{\rm diag}/(JL)=2/\sqrt{L}$, Eq.~(\ref{eq:sigma_width}) and $\sigma_{\rm can}$
at $U/J=0$ from Eq.~(\ref{sigmacanU0}).
At $U/J=1$, we use ED data, except for the $L=14$ and $20$ data points which were obtained from QMC.
}
\label{fig:sig_vsL}
\end{figure}

First, we compare $\sigma_{\rm can}$ and $\sigma_{\rm diag}$ for larger $L$ in Fig.~\ref{fig:sig_vsL} for $U=0$ and $U=J$ (for the latter up to $L=20$).
For the diagonal ensemble, $\sigma_\text{diag}/(JL) = 2/\sqrt{L}$ is an exact result. 
As expected, also the widths $\sigma_\text{can}/(JN)$ of the canonical ensemble approach a power-law decay with
$\sigma_\text{can}/(JL) \propto 1/\sqrt{L}$ for large $L$.
At $T=\infty$, as is the case for the quench to $U/J=0$, $\sigma_\text{can}$ can be computed exactly by combinatorially averaging over the position basis Fock states. For arbitrary $N$ and $L$ we obtain
\begin{equation} \label{sigmacanU0}
\frac{\sigma_\text{can}}{NJ} = \sqrt{1+\frac{N-1}{L+1}} \sqrt{\frac{2}{N}} \; ,
\end{equation}
which is plotted in Fig.~\ref{fig:sig_vsL}.
For $N=L$ and $L\gg 1$, it reduces to 
\begin{equation}
\sigma_\text{can}/(LJ) \approx 2/\sqrt{L} = \sigma_\text{diag}/(LJ)\,. 
\end{equation}
Therefore, in the limit $N\to \infty$, the prefactors of the power-law decays of $\sigma_\text{can}$ and $\sigma_\text{diag}$ become the same.
        This means that we expect our findings concerning the agreement of the diagonal and the canonical ensembles at small $U/J$ to persist
at large $L$.

\begin{figure}[t]
\includegraphics[width= 0.96\columnwidth]{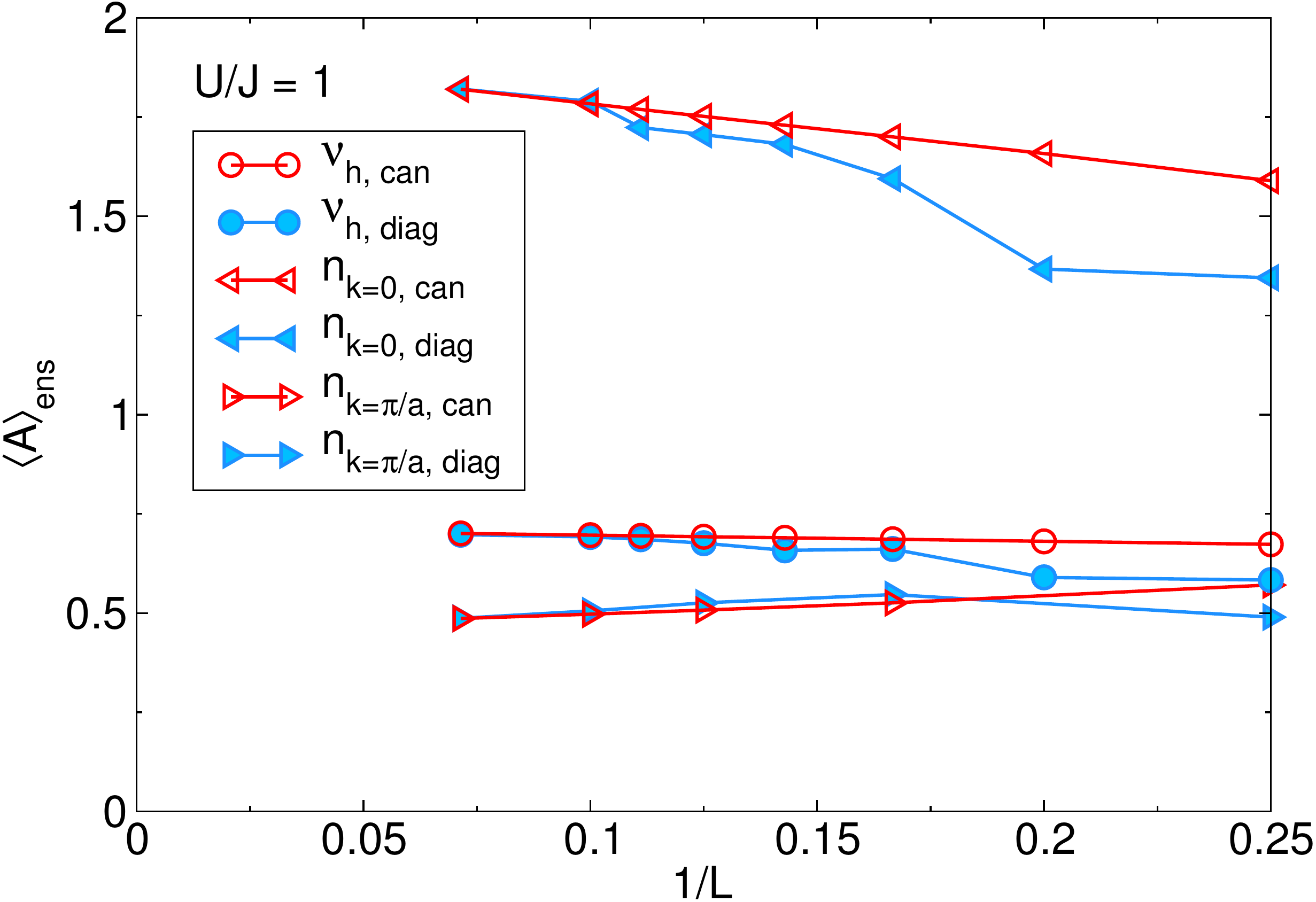}
\caption{(Color online)
Finite-size scaling of $\nu_h$, $n_{k=0}$, and $n_{k=\pi /a}$  in the diagonal and canonical ensemble
at $U=J$ using PBC (ED data, except for $L=14$, which is QMC data for the canonical ensemble).
}\label{fig:scalingU1}
\end{figure}

We next compute the expectation values of $\hat \nu_h$, $\hat n_{k=\pi/a}$ and $\hat n_{k=0}$ in the diagonal and the canonical ensemble for various system sizes
at $U=J$. These results are shown in Fig.~\ref{fig:scalingU1} and based on these data we conclude that the agreement between the canonical 
and diagonal ensemble survives for $L>10$. The scaling of $\sigma_{\rm can} \propto \sqrt{L}$ can in fact be understood
from the $L$-dependence of the specific heat. 

\begin{figure}[t]
\includegraphics[width= 0.96\columnwidth]{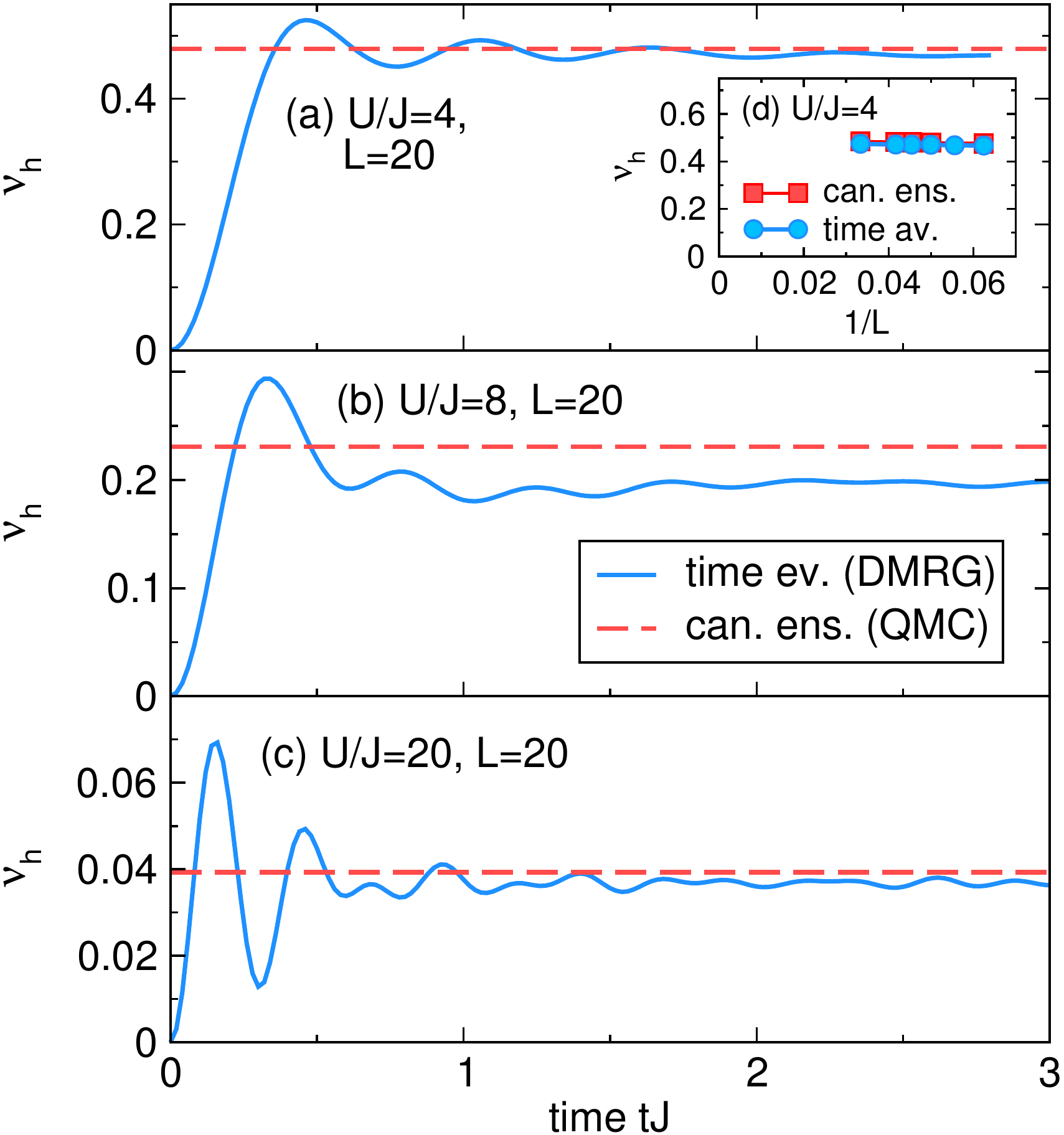}
\caption{(Color online)
Time evolution of $\nu_h$ (tDMRG, solid line) and canonical expectation value (QMC, dashed line)  for $L=20$, OBC, at (a) $U/J=4$, (b) $U/J=8$,
and (c) $U/J=20$. (d) Comparison of DMRG time averages for $t\in[2/J,t_{\rm max}]$ with  QMC data for the expectation value computed in the canonical ensemble as a function of inverse system size $1/L$ for $L=16,18, 20, 22, 24, 30$. Statistical errors in the QMC data are negligible and thus not shown.
}\label{fig:ens_obc}
\end{figure}

Figure~\ref{fig:ens_obc} shows tDMRG results for $\nu_h$ for $L=20$ and OBC, in comparison  with the canonical expectation values
computed with QMC. 
Obviously, for $U/J=4$, the time average agrees quite well with the QMC, whereas at $U/J=8$, the canonical
ensemble provides a poor approximation to the time average. For $U/J=4$, we computed time averages and canonical
expectation values for several larger systems shown in the inset of Fig.~\ref{fig:ens_obc}(a), 
with a very good agreement.
For $U/J=20$, the canonical ensemble is surprisingly close to the time average already at $L=20$,
similar to our expectation from the $L=10$ data (PBC) shown in Fig.~\ref{fig:tav}.

\subsection{Behavior for other initial states}

Finally, we mention that the similarity between expectation values in the canonical ensemble and the diagonal ensemble
is not restricted to pure Fock states in real space. We have verified that this behavior also emerges for
quenches from the ground state of the Bose-Hubbard model in the MI phase at finite values of $(U/J)_{\rm crit} < U/J<\infty$ (results not shown here).
Other possible initial  states that are product states in real space 
are those that have periodic arrays of doublons and empty sites with filling $n=1$, as studied in \cite{carleo12}, or the density-wave state
discussed in Refs.~\cite{cramer08,cramer08a,flesch08,trotzky12} with filling $n=0.5$.
Preliminary results for small systems show that micro-canonical and canonical ensemble agree well
with the time averages at small $U/J \leq 2$ for such initial states, while deviations between the 
thermal ensembles and the diagonal one become substantial for large $U/J>4$. This suggests
that the initial state being  the ground state of the pre-quench Hamiltonian 
could be another relevant ingredient in understanding the properties of the diagonal ensemble,
besides the initial state being a product state in real space.

%%%%%%%%%%%%%%%%%%%%%%%%%%%%%%%%%%%%%%%%%%%%%%%%%%%%%%%%%%%%%%%%%%%%%%%%%%%%%%%
\section{Sudden expansion experiment}
\label{sec:exp}

We now take advantage of the discussion of the diagonal ensemble presented in the paper and address some open questions emerging from recent experiments with ultracold atoms.
The purpose of this section is to provide an analytical interpretation of 
the numerical and experimental results for the expansion velocity in the sudden expansion of bosons in the
one-dimensional Bose-Hubbard model  in the large $U/J$ regime \cite{ronzheimer13}.
The expansion velocity is large at $U=0$ and at $U/J\gg 4$ and takes a minimum in the vicinity of 
$U=3J$ (see Fig.~\ref{fig:expansion}). The large expansion velocities at $U=0$ and $U/J=\infty$
are due to the ballistic expansion of free bosons and hard-core bosons in one dimension, respectively \cite{ronzheimer13}, while   
the reduction at intermediate $U/J$ is a consequence of the interaction quench \cite{vidmar13} that was performed in \cite{ronzheimer13}
simultaneously with the trap removal.  Due to the interaction quench, doublons and higher occupancies 
are dynamically generated, which was observed in the experimental data from \cite{ronzheimer13}.

We provide an analytical expression for the expansion velocity based on a two-component picture of 
ballistically expanding single atoms and inert doublons, valid for $U/J>10$.

\subsection{Fraction of doublons in the diagonal ensemble}
\label{subsec:fraction}

At small $L$ (small in terms of Eq.~(\ref{lcond})), $E$ sits in the lowest gap of the spectrum, see Fig.~\ref{fig:rho_U4_U10}(b).
The simplest possible approximation to the time-dependent wave-function can be taken for an effective two-level system,
\begin{equation}
\ket{\Phi(t)} = c_{\rm 0} e^{-iE^{(0)}t} \ket{\phi^{\rm (0)}} + c_1 e^{-iE^{(1)}t} \ket{\phi^{(1)}},
\end{equation}
where $\ket{\phi^{\rm (0)}}$ and $\ket{\phi^{(1)}}$ represent the ground state and a typical excited state in the first continuum of excited states, respectively.
We assume that these states can be expressed by the effective holon-doublon quadratic model~\cite{cheneau12,queisser14}.
In this case, $\hat \nu_h$ is a diagonal operator in the energy eigenbasis and therefore,  the fraction of doublons becomes a time-independent quantity,
$\tilde{\nu}_h(t) = \bra{\Phi(t)} \hat\nu_h \ket{\Phi(t)} = \braket{\tilde{\nu}_h}_{\rm diag} = \frac{2}{N} |c_1|^2$.
The coefficient $|c_1|^2$ can be obtained from  total energy conservation,
which gives
\begin{equation}
|c_1|^2 = \frac{-E_{\rm 0}}{U-6J} = \frac{4J^2}{U(U-6J)} N\,,
\end{equation}
where $E_0$ was obtained from  second-order perturbation theory.
The fraction of doublons in such an effective model is therefore equal to
\begin{equation} \label{doublons_diag}
\braket{\tilde{\nu}_h}_{\rm diag}= \frac{8}{\frac{U}{J} \left(\frac{U}{J} -6\right)}.
\end{equation}
The dashed line in Fig.~\ref{fig:tav}(a) represents $\braket{\tilde{\nu}_h}_{\rm diag}$ versus $U/J$, which agree very well with $\braket{\hat \nu_h}_{\rm diag}$ in the large $U/J$ limit.
Since all other excited states were neglected in this calculation, we expect that Eq.~(\ref{doublons_diag}) represents a lower bound for the exact $\braket{\hat \nu_h}_{\rm diag}$.

In addition, we also discuss the case where the number of particles is large, i.e., the conditions in Eq.~(\ref{ncond}) and Eq.~(\ref{lcond}) are fulfilled.
We assume that the ETH works in this regime, i.e., the diagonal and micro-canonical averages yield the same value, and that $\hat\nu_h$ is a diagonal operator in the energy eigenbasis.
Then we equate the quench energy with the excitation energy of a typical excited state with $m$ double occupancies (with respect to the ground state),
$\delta E = m(U-6J\eta)$,
where $m$ is an integer and $\eta \in [-1,1]$ takes into account the effective dispersion of holons and doublons.
Since we do not have any knowledge of $\eta$ (it depends on $N$ and $U/J$), we rather equal the quench energy to an effective level,
$\delta E = \tilde{m}(U-6J)$,
where $\tilde m$ represents a positive real number.
This induces some error into the calculation of $m$, which is, however, smaller than one if $U/J > 12$ (the full width of a single continuum of excitations, within the same $m$, is $12J$).
The fraction of double occupancies is then
\begin{equation} \label{nuh_largeN}
\nu_h \approx \frac{2\tilde m}{N} =  \frac{8}{\frac{U}{J} \left(\frac{U}{J} -6\right)},
\end{equation}
which is the same result as for $\braket{\tilde{\nu}_h}_{\rm diag}$ at small $N$, Eq.~(\ref{doublons_diag}).
This indicates that, as long as the fraction of doublons is concerned, the diagonal average barely depends on the number of particles.

\subsection{Expansion velocity in the large $U/J$ regime in a two-component picture}

We now apply our results of the previous section to 
 the sudden expansion of 1D bosons~\cite{ronzheimer13}, where the initial state was the same as in the present study.
The typical number of particles in a single tube in that experiment is $N\lesssim 80$.
Results in Eq.~(\ref{doublons_diag}) and Eq.~(\ref{nuh_largeN}), however, suggest that $\braket{\hat \nu_h}_{\rm diag}$ is not sensitive to the number of particles in the system.

\begin{figure}[bt]
\begin{center}
\includegraphics[width=\columnwidth]{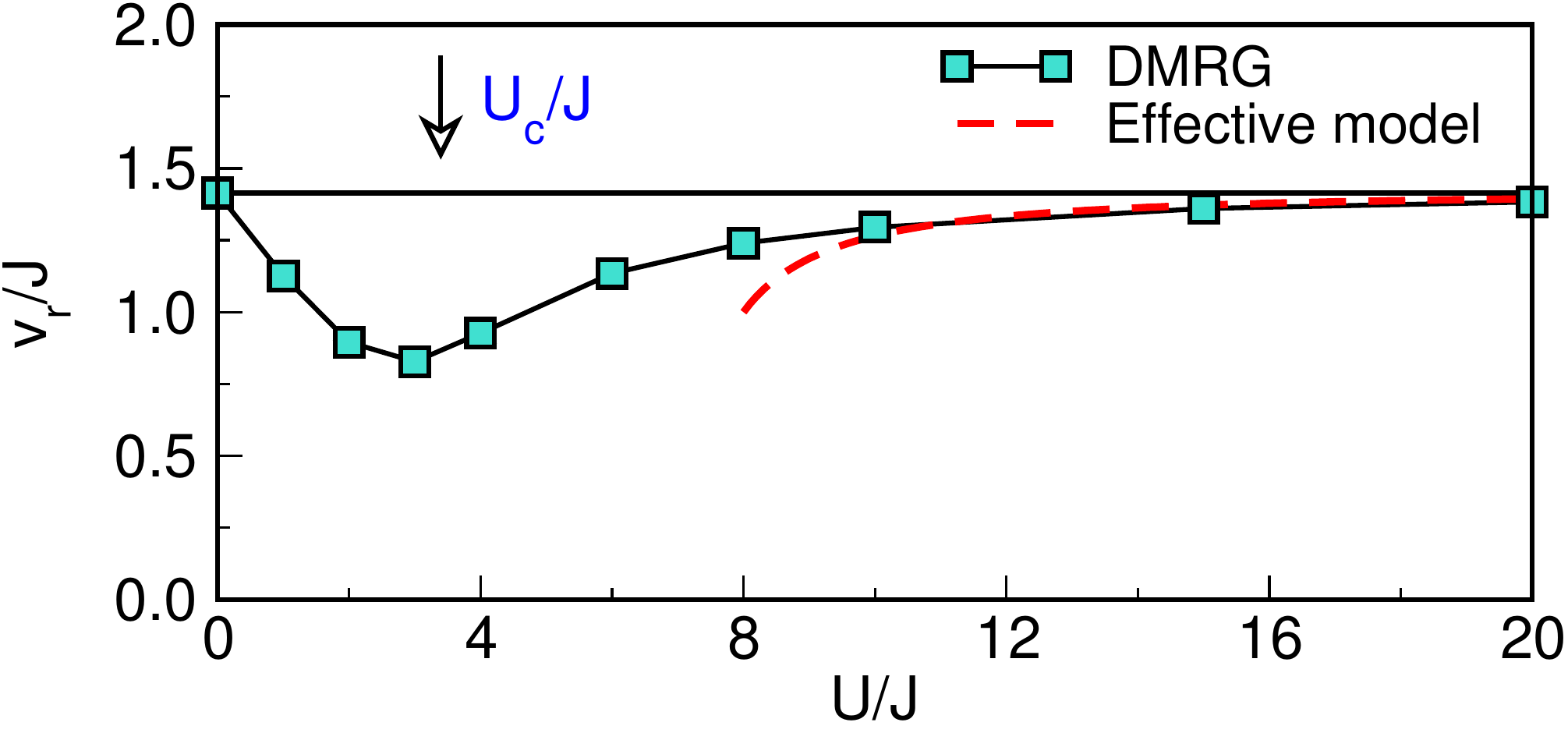}
\caption{
(Color online) Radial expansion velocity $v_{r}/J$ vs $U/J$.
Symbols: tDMRG data from Ref.~\cite{ronzheimer13}.
Dashed line: result from Eq.~(\ref{vr2}).
}
\label{fig:expansion}
\end{center}
\end{figure}

In the sudden expansion experiment, two quenches are performed:
(i) sudden removal of the trapping potential;
(ii) quench from infinite to finite $U/J$.
The time-dependent radius of the density distribution is defined as
\begin{equation}
R^2(t)  = \frac{1}{N} \sum_i \langle \hat n_i(t) \rangle (i-i_0)^2, \label{r_def}
\end{equation}
where $i_0$ represents the center of mass.
The corresponding {\it radial} velocity $v_{\mbox{\footnotesize r}}(t)$ is defined through the reduced radius $\tilde R(t)=\sqrt{ R^2(t) - R^2(0)}$ as
\begin{equation}
v_{\rm r}(t) = \frac{\partial \tilde R(t)}{\partial t}.
\end{equation}
We make the following assumptions, which are justified by the  numerical data from Refs.~\cite{ronzheimer13,vidmar13}:
(i) Doublons form very fast, i.e., on timescales for which the particles have not yet considerably expanded into an empty lattice. Accordingly, $\nu_h(t)$ approaches  $\braket{\hat \nu_h}_{\rm diag}$ rapidly as well.
(ii) Since $U/J$ is large, only a small fraction of doublons is formed. We do not take into account any higher occupancies, which are suppressed
at large $U/J$. 
(iii) Doublons, upon opening the trap,  undergo the quantum distillation process \cite{hm09,muth12,bolech12}, which makes them accumulate in the center of the lattice.
(iv) The doublons in the center of the lattice do not contribute to the radial expansion velocity. The latter is measured at large enough times such that the contribution of $R^2(0)$ can be neglected.

Following the last argument, we rewrite the radius such that
\begin{equation} \label{rnew}
R^2(t)  = \frac{N-2m}{N} \frac{1}{N-2m} \sum_i \langle \hat n_i(t) \rangle (i-i_0)^2.
\end{equation}
In this approach, there are $N-2m$ particles which yield the expansion velocity equivalent to the one of hard-core bosons, $v_{\rm r}^{\rm \tiny HCB}$, and the portion of atoms that are repulsively bound in  doublons gives $v_{\rm r}^{\rm d}=0$.
According to Eq.~(\ref{rnew}), the expansion velocity then reads as
\begin{equation}
v_{\rm r} = v_{\rm r}^{\rm \tiny HCB}  \sqrt{\frac{N-2m}{N}} = v_{\rm r}^{\rm \tiny HCB}  \sqrt{1-\nu_h}.
\end{equation}
This is a very convenient form of the expansion velocity since it requires only two input parameters:
(a) The expansion velocity of hard-core bosons, $v_{\rm r}^{\rm \tiny HCB} =\sqrt{2}J$~\cite{vidmar13};
(b) The fraction of doublons formed as a consequence of the additional quench from $U/J=\infty$ to a finite $U/J$.
Introducing the value of $\nu_h$ from Eq.~(\ref{nuh_largeN}), we can express $v_{\rm r}$ versus $U/J$ as
\begin{equation} \label{vr2}
v_{\rm r} = \sqrt{2} J\sqrt{1-\frac{8J^2}{U(U-6J)}}.
\end{equation}
The latter expression is plotted in Fig.~\ref{fig:expansion} as a dashed line.
It shows a very good agreement with tDMRG simulations for $U/J \gtrsim 10$, suggesting that the slow expansion
at large $U/J$ can be understood in a two-component picture of fast atoms and slow repulsively bound doublons.
Note that the role of doublons on the expansion of Bose and Fermi gases in 1D and 2D has also been discussed in Refs.~\cite{hm09,kajala11,jreissaty13},
and the scaling of $v_{\rm r}(U/J)$ from Eq.~(\ref{vr2}) has been recently applied to describe the expansion of two bosons in Ref.~\cite{boschi14}.

\section{Summary and Discussion}
\label{sec:sum}

In this work we studied the relaxation and thermalization dynamics in the one-dimensional Bose-Hubbard model,
starting from a product state with exactly one boson per site. This particular state was realized in recent experiments \cite{cheneau12,ronzheimer13}. We focused our attention on the observables relevant for Ref.~\cite{ronzheimer13}, namely the fraction $\nu_h$ of atoms on multiply occupied sites and the MDF.

Our main result is that both the micro-canonical and the canonical ensemble describe well the time averages 
of the fraction of atoms on multiply occupied sites and the MDF  in the regime $0<U/J\lesssim 5$, already on systems of $L\gtrsim 10$.
This  is consistent with the  observation from Refs.~\cite{buchleitner,kollath10} of chaotic properties in the spectrum of the Bose-Hubbard model at these values
of $U/J$.
This suggests that one can assume at least local equilibrium with temperatures extracted from the 
canonical ensemble to faithfully describe the state of the cloud after just a short transient dynamics of the order of $t\sim 2/J$.
Potentially longer time scales involved with global thermalization of energy fluctuation patterns \cite{lux13} or a possible transition from
prethermalization to thermalization (see also \cite{queisser14,essler14}) 
are not captured on the system sizes and for the observables studied here, yet deserve future investigations.

 In the small $U/J$ regime, we nevertheless observe cases
in which the micro-canonical ensemble deviates from the time averages  for the MDF. These instances as well as the overall
good agreement between the micro-canonical  and the diagonal ensemble can be understood by inspecting the 
criteria for the validity of the ETH, namely narrow and smooth distributions of eigenstate expectation values of observables
and a sufficient narrowness in energy of the diagonal ensemble, defined by the initial state.
In order to analyze the numerical data with respect to the ETH, we suggest that one should take into account both criteria
by studying the fluctuations of $\langle \alpha | \hat A | \alpha \rangle$ over a window in energy around the system's  energy  given by the width defined by
the initial state. This   perspective on the data analysis is complementary  to other studies \cite{santos10,steinigeweg14,beugeling14}, which emphasize the initial state independence
of the ETH concept, while here, we are interested in elucidating at which $L$ observables for a very specific quench 
appear to be thermalized.
Note that in the example studied here, the issues with the  micro-canonical ensemble can be understood in terms of obvious finite-size effects
that do, however, not contradict  ETH becoming applicable for sufficiently large system sizes.

Very interestingly, the canonical ensemble describes the time averages  of all observables studied here at least
as well as the micro-canonical ensemble, but even works where the micro-canonical ensemble fails on small systems.
In the limit of small $U/J$, we trace this back to two aspects: First, the canonical temperature is infinite at $U=0$
and therefore the canonical energy distribution is perfectly Gaussian and in fact very similar to the diagonal energy 
distribution. Second, it appears that initial states that are product states in real space lead to a particular form of 
the diagonal ensemble that eventually implicates its similarity to the canonical ensemble. This latter observation has also 
been stated  in Ref.~\cite{rigol11} for initial product states and quenches to integrable  models. Properties of the diagonal ensemble can neatly be studied
by computing the short-time dynamics of the fidelity, without any need for a full diagonalization \cite{roux10,torres-herrera14,torres-herrera14a} (see Appendix~\ref{sec:fidel}).

Finally, we used our numerical results for the time-average of $\nu_h$ as a function of $U/J$ to back up
an approximate calculation of the expansion velocity in a sudden expansion starting from the same initial state 
as in our work. We assume that for large $U/J$, only doublons are dynamically generated, which, due to quantum
distillation processes \cite{hm09,bolech12,muth12} accumulate  in the center of the trap. Thus, in this regime the expansion 
velocity can be understood as due to
a ballistic component minus the inert portion that is repulsively bound in the doublons. This yields an 
estimate of the expansion velocity for $U/J\gtrsim 10$, which quantitatively agrees with numerical data from Ref.~\cite{ronzheimer13},
and corroborates the qualitative picture put forward in Ref.~\cite{ronzheimer13}.

\emph{Acknowledgments} --- We are indebted to J. Gemmer, S. Kehrein, M. Rigol, G. Roux,  and U. Schneider for very fruitful discussions.
We thank M. Rigol and L. Santos for their comments on a previous version of the manuscript.
We acknowledge support from the Deutsche Forschungsgemeinschaft (DFG) through FOR 801. 
L.V. is supported by the Alexander von Humboldt Foundation.

\appendix

\section{Time evolution of the fidelity}
\label{sec:fidel}
The fidelity $F(t)$ is defined \cite{peres84,roux09,torres-herrera14, torres-herrera14a} as the probability to find the initial state 
$ \ket{\psi_\text{in}} =\ket{\psi(t=0)} = \sum_\alpha c_\alpha \ket{\alpha}$ at a given time $t$,
\begin{align}
F(t) &= |\braket{\psi_\text{in} | \psi(t)}|^2 
= |\bra{\psi_\text{in}} e^{-iHt} \ket{\psi_\text{in}}|^2 \nonumber \\
&= \left|\sum_\alpha |c_\alpha|^2 e^{-iE_\alpha t}\right|^2 \ .
\end{align}
Using our definition Eq.~(\ref{eq:diag_en_distr}) of the energy distribution of the diagonal ensemble, $\rho_\text{diag}(E) = \sum_{\alpha} |c_{\alpha}|^2 \delta(E-E_{\alpha})$, which is nothing but the sum of the initial state's eigenstate occupations times the density of states, we can rewrite this as
\begin{equation}
F(t) = \left| \int_{-\infty}^{\infty} \rho_\text{diag}(E)e^{-iEt}dE \right|^{2} \, ,
\label{eq:fid_FT}
\end{equation}
the fidelity thus being given by the modulus squared of the Fourier transform of $\rho_\text{diag}(E)$.

\begin{figure}[t]
        \includegraphics[width=\columnwidth]{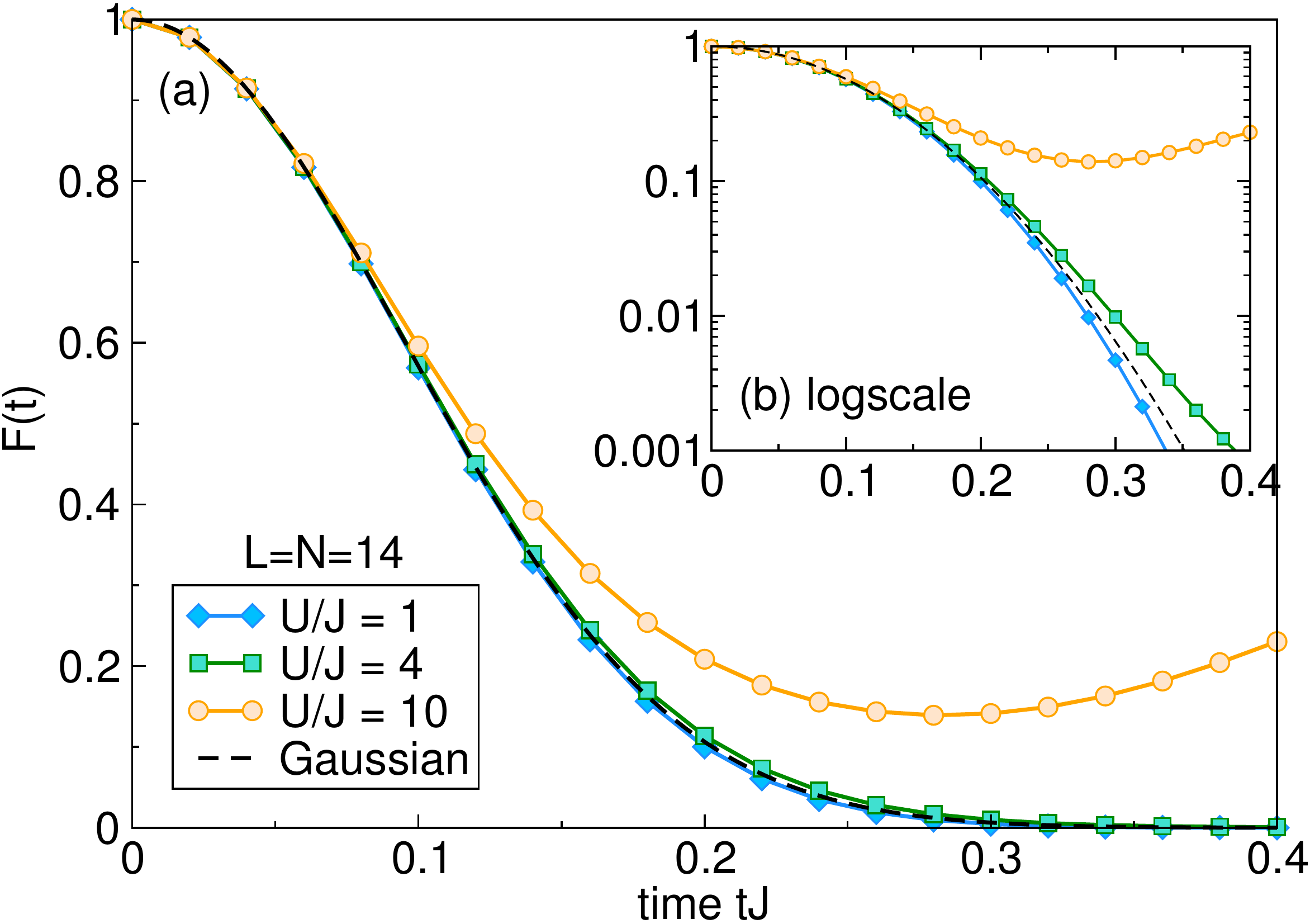}
        \caption{
        {(a)} Time evolution of the fidelity in the transient regime after the quench from $U/J=\infty$ to $U/J=1,4,10$ for a system with $L=14$ and PBC.
        {(b)} Time evolution as in (a), but with a logarithmic scale for the fidelity.
        }\label{fig:fid}
\end{figure}

Here we are interested in the short-time dynamics of $F(t)$ (see also the discussion in Refs.~\cite{torres-herrera14, torres-herrera14a}) and the direct relation to $\rho_\text{diag}(E)$ through Eq. (\ref{eq:fid_FT}).
For example, if $\rho_\text{diag}(E)$ can sufficiently well be approximated by a normal distribution of width $\sigma_\text{diag}^2 = \bra{\psi_\text{in}} (H- E)^2 \ket{\psi_\text{in}} = \bra{\psi_\text{in}} H^2 \ket{\psi_\text{in}} - \bra{\psi_\text{in}} H \ket{\psi_\text{in}}^2 = \sum_\alpha |c_\alpha|^2 (E_\alpha -  E)^2$, centered around the system's energy $E = \bra{\psi_\text{in}} H \ket{\psi_\text{in}}$, then the time evolution of the fidelity will  just be the absolute square of the Fourier transform of a Gaussian, i.e., another Gaussian,
\begin{align}
F(t) &\approx \left|\frac{1}{\sigma_\text{diag} \sqrt{2\pi}} \int_{-\infty}^\infty e^{-(\epsilon- E)^2/(2\sigma_\text{diag}^2)} e^{-i\epsilon t} d\epsilon \right|^2 \nonumber\\
&=\left| e^{-\sigma_\text{diag}^2 t^2/2} e^{-iE  t} \right|^2 = e^{-\sigma_\text{diag}^2 t^2} \ .
\label{eq:fid}
\end{align}
By the same reasoning, a Lorentzian distribution of energies $\rho_\text{diag}(E)$ corresponds to an exponential decay of $F(t)$.

For $\ket{\psi_\text{in}}=\prod_{i=1}^{L} a^{\dag}_{i} \ket{0}$, we have $E=0$ and $\sigma_\text{diag} = 2J\sqrt{L}$, independent of $U$, since $\bra{\psi_\text{in}} H \ket{\psi_\text{in}} = 0$ and $\bra{\psi_\text{in}} H^2 \ket{\psi_\text{in}} = \braket{H\psi_\text{in}| H\psi_\text{in}} = (-J\sqrt{2})^{2} \cdot 2L = 4J^{2}L$ --- each of the $2L$ hopping terms uniquely creating one double occupancy and one empty site, the diagonal terms yielding zero.

Typical results for the short-time evolution of the fidelity are presented in Fig.~\ref{fig:fid}.
As expected, the time evolution in the transient regime is properly reproduced by the Gaussian given in Eq.~(\ref{eq:fid})  (compare Ref.~\cite{torres-herrera14a}, where the same behavior was observed for other examples) as long as the energy distribution $\rho_\text{diag}(E)$ (see Figs.~\ref{fig:DoSU1} and~\ref{fig:rho_U4_U10}) can reasonably well be approximated by a normal distribution ($U/J=1$ and $4$).
Deviations become significant already at times clearly below the transient time scale if this is not the case ($U/J=10$).

The results for the specific initial state discussed in Fig.~\ref{fig:fid} suggest that the time evolution of the fidelity in the transient regime does indeed constitute an effective measure for the energy distribution $\rho_\text{diag}(E)$ of the initial state, with the advantage that the time evolution is accessible through Krylov methods or tDMRG, which allows studying larger system sizes than what is accessible to 
complete diagonalization.

\section{Convergence of the time-dependent DMRG data}
\label{sec:error}

\begin{figure}[t]
        \includegraphics[width=\columnwidth]{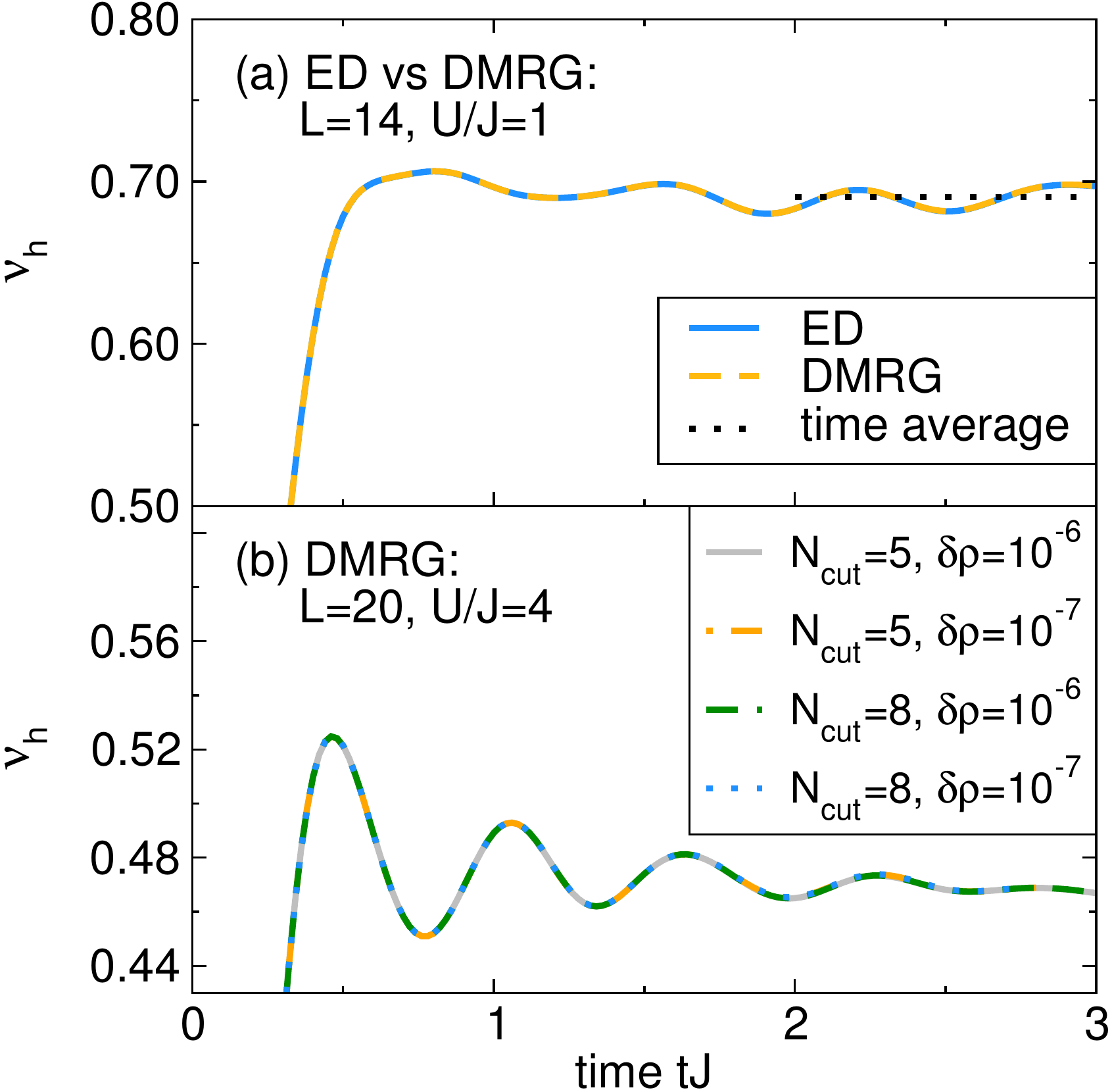}
        \caption{Time evolution of the fraction $\nu_h$ of atoms on multiply occupied sites for OBC.
        {(a)} $L=14$, $U/J=1$ calculated using ED and tDMRG ($N_{\rm cut}=8$, $\delta\rho=10^{-7}$), with the time average being taken over the tDMRG data for $tJ\in[2,3]$.
        {(b)} $L=20$, $U/J=4$ calculated using tDMRG with the maximal number of bosons per site $N_{\rm cut}=5, 8$ and with maximal discarded weights of $\delta\rho=10^{-6}$ and $\delta\rho=10^{-7}$. 
        The time step is in all cases chosen as $0.02/J$.
        %The data shown is converged for all parameters.
        }\label{fig:con}
\end{figure}

We finally comment on the convergence of the tDMRG method that we use to calculate the relaxation dynamics.
In Fig.~\ref{fig:con}(a) we compare the results of $\nu_h$ obtained by tDMRG to the exact time evolution at $U/J=1$, using OBC.
The data show perfect agreement.
For all the results obtained by tDMRG, we ensure that the discarded weight $\delta \rho$ is small enough and the local bosonic cut-off $N_{\rm cut}$ large enough to get converged results.
In Fig.~\ref{fig:con}(b) we show  examples for  different choices of $N_{\rm cut}$ and $\delta \rho$, which all lead to virtually indistinguishable curves.

%%%%%%%%%%%%%%%%%%%%%%%%%%%%%%%%%%%%%%%%   Bibliography

%\bibliographystyle{apsrev}
\bibliography{references}

\end{document}